\documentclass[12pt,preprint]{aastex}
\usepackage{graphics}

\begin{document}
\slugcomment{Submitted to AJ @ \today}

\title{Optical studies of ultra-luminous X-ray sources in nearby galaxies}

\author{Ji-Feng Liu, Joel N. Bregman, Patrick Seitzer, and Jimmy Irwin}
\affil{Astronomy Department, University of Michigan, MI 48109}

\begin{abstract}

Optical studies of ultra-luminous X-ray sources is an essential step in understanding the nature of these enigmatic sources, and in this paper we report our studies of five ULXs in NGC4559, NGC5194, NGC1313 and NGC628 observed with the Hubble Space Telescope and the 6.4 Magellan/Baade telescope.  The ULX in NGC4559 is identified with four blue and red supergiants within a $0\farcs2$ error circle, in a star forming region that is younger than $10^{7.4}$ years.  ULX-3 in NGC5194 is located on the rim of a star cluster with a few faint stars within the $0\farcs3$ error circle, though the secondary in this system is expected to be a low mass ($0.3M_\odot$) star below detection limits given its two hour period.  ULX-5 in NGC5194 is located on a spiral arm with seven objects within the $0\farcs8$ error circle, which can be improved with future observations.  Both ULXs in NGC5194 are in regions younger than $10^{7.8}$ years.  The ULXs in NGC1313 and NGC628 was observed with the Baade telescope.  The ULX in NGC1313 is identified with one object with R-I color bluer than the bluest stellar objects, indicative of non-thermal emission in the R band, or dramatic variability between two observations, or errors in measurements.  The ULX in NGC628 is located in a bubble nebula and identified with a few extended objects that are probably young stellar clusters.  Future observations with better spatial resolutions are needed to clarify the optical counterparts and their nature for the ULXs in NGC1313 and NGC628.

\end{abstract}

\keywords{X-rays: binaries -- X-rays: galaxies -- Optical: binaries}

\section{INTRODUCTION}

Ultra-luminous X-ray sources (ULXs) are extra-nuclear sources with luminosities
in the range of $ 10^{39}-10^{41}$ erg/sec in other galaxies, and have been
observed by Einstein (e.g., Fabiano et al. 1989), ROSAT (e.g., Colbert 1999),
ASCA (e.g., Makishima et al. 2001), recently by XMM-Newton and Chandra
Observatory in many galaxies (e.g., Kilgard et al. 2002; Swartz et al. 2004).
As compared to the cases of the X-ray binaries in our Galaxy, which are powered
by accretion onto neutron stars or stellar mass black holes and have
luminosities of $ 10^{33}-10^{38}$ erg/sec, the luminosities of ULXs require
accreting black holes of masses  $10^3$ -- $10^4$ $M_\odot$ if they emit at
$10^{-2}$ of the Eddington luminosity, typical of Galactic X-ray binaries.
Such intermediate mass black holes (IMBHs), if they exist, bridge up the gap between
stellar mass black holes and supermassive black holes of $10^6$ -- $10^9
M_\odot$ in the center of galaxies.
However, such IMBHs are not predicted to be the products of ordinary stellar
evolution models, and it is still in debate whether they can form in dense
stellar fields via runaway stellar collisions with seed BHs of a few hundred $M_\odot$
(e.g., Portegies Zwart 2002).
Alternatively, these sources can be stellar mass black holes, for which the
formation is well established in theory and observation.
Special mechanisms are required for these stellar mass black holes to emit at
super-Eddington luminosities.
Such a mechanism could be extreme relativistic beaming (Georganopoulos et al.
2002), mild geometric beaming (e.g., King et al. 2001; Abramowics et al. 1978),
or the photon-bubble instability in a radiation pressure dominated accretion
disk, which leads to truly super-Eddington luminosities (Begelman 2002).

The ultimate goal in ULX studies is to determine the mass of the primary of ULX
systems.
While the primary mass can be estimated through the X-ray spectra (e.g.  Miller
et al. 2003) or possible quasi-periodic oscillations (e.g. Strohmayer et al.
2003), the oldest yet most secure way is to measure the mass function through
the secondary mass and the orbital parameters such as the velocities, the
period, and/or the orbit size, which could be done through spectroscopic
observations of the optical secondary.
The first step of this method is to identify the secondary in optical, which is
challenging because ULXs are usually located in star forming regions where
star density is very high and there could be a few tens of objects even within
a small error circle of $1^{\prime\prime}$ radius.
To identify a ULX to one unique counterpart, one needs its X-ray position
accurate to a few tenth of arcseconds. Such positional accuracy can only be
obtained with Chandra observations, which have uncertainties of $\sim0\farcs6$,
and can be further improved with relative astrometry.
One also needs to be able to resolve the bright objects in the error circle,
which can be done for star forming regions only in nearby galaxies given the
limitation on the spatial resolution of the telescopes of our time.

Optical studies of ULXs and their environments have been carried out with both
large ground-based telescopes and the Hubble Space Telescope.
Pakull \& Mirioni (2002) surveyed fifteen ULXs in eleven nearby galaxies with
ground based telescopes, and found HII regions, (bubble) nebulae, and X-ray
ionized nebulae (XIN) within the error circles of their ROSAT HRI positions.
Kaaret et al. (2004) observed the XIN at the location of the ULX in Holmberg II
with HST/ACS, and found a bright, point-like counterpart for which the
magnitude and color are consistent with a star between O4V and B3 Ib, or
reprocessed emission from an X-ray illuminated accretion disk.
Zampieri et al. (2004) found for the ULX in NGC1313 a faint optical counterpart
within the surrounding bubble nebula with ESO 3.6m telescope. 
Cropper et al. (2004) found multiple objects on HST/WFPC2 images within the
$0\farcs7$ error circle around the nominal Chandra position for the ULX in
NGC4559.

We have studied a sample of ULXs in nearby galaxies with HST and the
Magellan/Baade 6.4-meter telescope.
This has led to the successful identifications of the ULX in NGC3031 (Liu et al.
2002) and the ULX in NGC5204 (Liu et al. 2004).
The ULX in NGC3031 is identified to a unique point-like object on the WFPC2
images, for which the magnitudes and colors are consistent with those for an
O8V star that is presumably the optical secondary in the ULX system. Unlike the
high-mass X-ray binary (HMXB) systems in our Galaxy that are accreting mass through
stellar wind, the O8V star in this system may be filling its Roche Lobe and
accreting mass onto the primary with much higher accretion efficiency that
leads to its high X-ray luminosity.
For the ULX in NGC5204, we have found a unique optical counterpart with the HST
ACS/HRC images and, from its HST STIS spectrum, determined the secondary to be
a B0 Ib supergiant which is presumably overflowing its Roche Lobe and accreting
materials onto the primary black hole. 
There is a NV emission line at 1240\AA \  in its spectrum suggestive of an
accretion disk, since the high ionization state of NV cannot occur in a B0 Ib
star, but is easily achievable in an accretion disk and its corona as observed
in some low mass X-ray binaries (LMXBs) such as LMC X-3 and Hercules X-1. 

In this paper we report our results on the rest ULXs in our sample.
Among these, one ULX in NGC4559 is identified to four neighboring objects in
HST/WFPC2 images with much improved astrometry,  two ULXs on the spiral arms in
NGC5194 show multiple objects in the error circles of Chandra positions, with
one ULX located on the rim of a star forming knot.
The ULXs in NGC1313 and NGC628 studied with the Baade telescope are also
reported here.


\section{DATA ANALYSIS AND RESULTS}

Optical studies of ULXs in star forming regions require the combination of very
accurate X-ray and optical positions and high spatial resolution of the optical
images.
For accurate X-ray positions, we utilize the Chandra ACIS observations, which
have a typical accuracy of $0\farcs6$ for absolute positions (Aldcroft et al.
2000).
The ACIS data were analyzed with CIAO 2.3 and CALDB 2.12, and WAVDETECT was run
to detect discrete sources on ACIS-S chips.
The X-ray luminosities for ULXs were estimated from their count rates by
assuming a power-law spectrum in 0.3--8 KeV with a photon index of 1.7,
consistent with the average photon index for ULXs in nearby galaxies (Swartz et
al. 2004).

The Hubble Space Telescope provides the best spatial resolution for optical
observations.
HST images in three filters of F450W (B),
F555W (V) and F814W (I) were taken for a sample of ULXs in nearby galaxies for
our HST program (GO 9073) with the Wide Field and Planetary Camera (WFPC2).
The pixel size for the Planetary Camera chip is $0\farcs05$, and $0\farcs1$ for
the three Wide Field chips. 
We used HSTphot (Dolphin 2000) to perform PSF fitting photometry. 
%
%
The absolute accuracy of the positions obtained from WFPC2 images is typically
$0\farcs5$ R.M.S. in each coordinate (HST data handbook).
The relative positions have higher accuracy, which is better than
$0\farcs005$ for targets contained on one chip, and
$0\farcs1$ for targets on different chips.

A direct overlay of the X-ray position on the HST WFPC2 image would yield an
error circle of $0\farcs8$, unable to identify the ULX with a unique object in
dense stellar fields.
However, the error circle can be greatly reduced when one or more nearby X-ray
sources are identified with objects on the same optical image, so that the ULX
can be registered on the optical image relative to the counterparts for the
nearby X-ray sources. 
This technique of relative astrometry involves the accuracies of relative
positions for objects on the X-ray and optical images which are better than the
accuracies of absolute positions, and usually leads to error circles of a few
tenths of arcseconds, small enough to enclose only one object in moderately
dense stellar fields.
This technique has been successfully applied for the ULXs in NGC3031 and
NGC5204, each of which is identified with a unique point-like object within
the much reduced error circle ($\sim0\farcs2$).
The technique is also used to improve the positional accuracies on the optical
images for the ULXs reported in this paper.


While limited by the seeing, the observations with large ground based
telescopes still play a role in identifying the optical counterparts for ULXs,
especially for those in less dense stellar fields.
We observed a sample of ULXs in nearby galaxies with the Inamori Magellan Areal
Camera and Spectrograph (IMACS) mounted on the 6.4-meter Magellan/Baade
telescope in November 2003, and achieved a seeing of $\sim0\farcs6$.
The IMACS images of 2X4 mosaic CCDs have a $15^\prime\times15^\prime$ field of
view with a pixel scale of $0\farcs11$/pixel.
The large field of view of the IMACS images is advantageous to relative
astrometry, since more X-ray sources with optical counterparts would be found
near the ULX in a larger image. This expectation is verified in the case of
the ULX in NGC628.
The photometry of our observations is calibrated with short exposures of the
photometric standard field
L98\footnote{http://cadcwww.dao.nrc.ca/cadcbin/wdb/astrocat/stetson/query/L98}.

In the following, we report the observations and analysis on the five ULXs in
NGC4559, NGC5194, NGC1313 and NGC628.


\subsection{A ULX in NGC4559}

NGC4559 is a Scd spiral galaxy with a distance reported from 9.7 Mpc
(Tully 1988) to 5.8 Mpc (Tully 1992).  X-ray observations with ROSAT, XMM, and
Chandra observatories have revealed two ULXs in the galaxy. One ULX is close to
but not the nucleus, another is $2^\prime$ south of the nucleus and located on
an outer spiral arm. Here we study the latter.

This ULX have been observed with Chandra three times in January 2001
(Observation ID 2026, 9 kiloseconds), June 2001 (ObsID 2027, 10
kiloseconds), and March 2002 (ObsID 2686, 3 kiloseconds).
The average position for the ULX from three Chandra observations is
R.A.=13h35m51.7s, Decl.=27d56m04.7s.
Placed at a distance of 5.8 (9.7) Mpc, the ULX shows an X-ray luminosity of
$9.2\times10^{39}$ ($2.6\times10^{40}$) erg/sec in June 2001, making it one of
the most extreme ULXs. Its luminosity dropped to $4.8\times10^{39}$
($1.3\times10^{40}$) erg/sec in March 2002, exhibiting a dramatic drop of 50\%
in nine months.
Cropper et al. (2004) studied the spectra in Chandra observations and in an XMM
observation, and found that all spectra show a soft component that can be
fitted with a disk temperature of $\sim0.12$ KeV, suggestive of a black hole of
$M_\bullet \ge 10^3 M_\odot$ if interpreted as the inner disk temperature of
the accretion disk around the black hole.
For timing analysis, they computed the power density spectrum for the ULX, and
found a break in the power-law fit at $f_b=28$ mHz, which indicates the black
hole mass $M_\bullet =$ 40 (1300) $M_\odot$ if $f_b$ is taken as the lower
(higher) low frequency break in the $M_\bullet$--$f_b$ scaling relation as
demonstrated in AGNs and X-ray binaries such as Cygnus X-1 (Belloni \& Hasinger
1990).


The ULX has been observed with HST/WFPC2 in filters F450W (B), F555W (V) and
F814W (I).
A direct comparison of the X-ray images with the WFPC2 images yields five
bright objects and many more faint objects within the $0\farcs7$ error circle
around the nominal Chandra position of the ULX (Cropper et al. 2004).
Fortunately, relative astrometry can be done to improve the positional accuracy
due to two nearby X-ray sources (X4 and X5) that show optical counterparts on
the WFPC2 images.
The ULX and X4 were detected in the second observation. X4 falls on the WF2
chip, and is identified with an extended object that is an elliptical galaxy or
an edge-on spiral galaxy in the background.
The ULX and X5 were detected in the third observation. X5 falls on the WF3
chip, and is identified with a bright point-like object that may be a
foreground star.
The X-ray positions and optical positions for X4 and X5, and derived optical
positions for the ULX are listed in Table 1.
The error circles around the corrected ULX positions derived from X4 and X5
have radii of $\sim0\farcs3$, and are plotted in Figure 1.
The two error circles partly overlap each other, and in the overlapping region is one
point-like object.
Another three point-like objects are within the error circle derived from X5. 
In total, four point-like objects are identified as candidates for the optical
counterpart of the ULX, and the rest in the $0\farcs7$ nominal error circle
adopted by Cropper et al. (2004) are excluded.
The magnitudes for the four candidates are calculated with HSTphot and listed
in Table 2.


The four candidates form a close association, and lie between two young star
clusters in a star forming complex on the outer edge of NGC4559. 
The star forming complex is isolated from other star forming regions/knots on
the inner spiral arms, and is probably triggered by a recent plunge of a nearby
dwarf galaxy through the disk (Soria et al. 2004).
The bright stars form two distinct groups on the color-magnitude diagram, one
of blue OB stars, and one of red supergiants (Figure 2).
The ages can be estimated by comparing the colors and magnitudes of the stars
to isochrones, for which we use the isochrones from the Geneva stellar models
(Lejeune \& Schaerer 2001) with Z=0.020 given the low metal abundance of the
environments (Soria et al. 2004).
As seen from Figure 2, the red supergiants span the age range of
$10^{7.4}$--$10^{7.6}$ ($10^{7.2}$--$10^{7.4}$) years if a distance of 5.8
(9.7) Mpc is adopted.  This is consistent with the age estimate of $\le$30 Myrs
by Soria et al. using a distance of 9.7 Mpc (2004).
The bright blue stars are younger than $10^{7.2}$ ($10^{7}$) years.
For the faint stars, the large errors in their colors prevents an estimate of
the ages.
Among the four candidates, three less brighter candidates are evolved red
supergiants in the age range of 25--40 (15--25) Myrs, while the brightest one
is a blue star/supergiant that is leaving the main sequence, with an age of
$\le$15 ($\le10$) Myrs.


The spectral type and luminosity class, thus the mass and size of a stellar
object can be inferred from comparing its absolute magnitudes and colors to
those of different MK types.
Here the extinction is corrected for the Galactic HI column density of
$1.5\times10^{20}$ cm$^{-2}$, which corresponds to $A_B$=0.106 mag, $A_V$=0.080
mag, and $A_I$=0.049 mag if we adopt the relation $n_H = 5.8\times10^{21}\times
E(B-V)$ (Bohlin, Savage, \& Drake 1978) and $R_V=3.1$.
The extinction intrinsic to NGC4559, though may be larger than the Galactic
extinction, is neglected since they must be small with $E(B-V)<0.1$ mag, or
else the blue OB population would be significantly bluer than stellar models
can predict as seen in Figure 2.
The color $B-V$ and the absolute magnitude $M_V$ for the four candidates (Table
2) are compared to those of  different MK types as tabulated in Schmidt-Kaler
(1982).
If we adopt a distance of 5.8 Mpc ($\mu$=28.82 mag), the candidate C-1 is
consistent with a B2 Ib supergiant ($M_V$=-5.7, $B-V$=-0.18), C-2 is consistent
with an F8 Ib supergiant ($M_V$=-5.1, $B-V$=0.56), C-3 is consistent with a G5
Ib supergiant ($M_V$=-4.6, $B-V$=1.00), and C-4 lies between F2 Ib ($M_V$=-5.1,
$B-V$=0.23) and F2 II ($M_V$=-2.4, $B-V$=0.30).

The candidates will be identified to different MK types if we adopt a different
distance thus different absolute magnitudes for the candidates.
For a distance of 9.7 Mpc ($\mu$=29.93 mag), C-1 is most consistent
with a B1 Ia supergiant ($M_V$=-6.9, $B-V$=-0.19).
For C-2, its color is consistent with an F8 Ib or F8 Iab supergiant
($M_V$=-6.5, $B-V$=0.56), while its $M_V$ is brighter than F8 Ib by 0.9 mag,
and dimmer than F8 Iab by 0.5 mag.
For C-3, its color is consistent with a G5 Ib or G5 Iab ($M_V$=-6.2,
$B-V$=1.02) supergiant, while its $M_V$ is brighter than G5 Ib by 0.5 mag, and
dimmer than G5 Iab by 0.6 mag.
For C-4, the color lie between an F2 Ib supergiant and an F5 Ib
supergiant ($M_V$=-5.1, $B-V$=0.33), while it is brighter than such supergiants
by 0.2 mag.
%


The above inferred spectral types are subject to errors due to uncertainties of
the distance and the intrinsic extinction.
For example, if the total extinction amounts to $E(B-V)$=0.1 mag (thus
$A_B$=0.408 mag and $A_V$=0.308 mag for $R_V$=3.1), C-1 would have
$M_V$=-5.97 and $B-V$=-0.26 at a distance of 5.8 Mpc, consistent with a
supergiant between B0 Ib ($M_V$=-6.1, $B-V$=-0.24) and B0 II ($M_V$=-5.7,
$B-V$=-0.29).
A more robust way to classify the candidates to MK spectral types is to take
spectra with low or medium resolutions that can reveal the spectral shape and
even line features.
The mass and size of the stars can be inferred empirically once the spectral
types are known, and the mass of the primary can be calculated if one star is
identified as the secondary with improved astrometry and the orbital velocities
is known with further spectroscopic observations.


\subsection{Two ULXs in NGC5194}

NGC5194 (M51) is a grand-design Sbc galaxy at a distance of 7.7 Mpc (Tully 1988) which is
interacting with its companion NGC5195.
Six ULXs have been discovered in M51 with ROSAT HRI observations (Liu \&
Bregman 2005), greatly exceeding the average of $0.72\pm0.11$ ULXs per spiral
galaxy as discovered in a recent HRI survey of ULXs in nearby galaxies (Liu,
Bregman \& Irwin, 2004).
All ULXs are located on the spiral arms, and here we study two of them, the
ULX-3 and ULX-5 as designated in Liu \& Bregman (2005).


ULX-3 has been observed with Chandra three times in January 2000 (ObsID 414, 1
kilosecond), June 2000 (ObsID 354, 15 kiloseconds), and in June 2001 (ObsID
1622, 27 kiloseconds). 
The average X-ray position for ULX-3 from the last two observations is
R.A.=13h30m01.0s, Decl.=47d13m44.0s.
Some interesting properties for this ULX has been revealed from the Chandra
observations (Liu et al. 2002).
In one year between the last two observations, the luminosity dropped
dramatically from $\sim2\times10^{39}$ erg/sec to $\sim8\times10^{37}$ erg/sec,
4\% of the original level.
Along with the dramatic luminosity decrease, the X-ray spectra changed from a
hard/high state to a soft/low state, in contradiction  to the soft/high to
hard/low transition commonly seen in Galactic LMXBs.
Most remarkably, a two hour periodic variation was revealed in the four hour
observation in June 2000, which makes ULX-3 one of the few that have periods.
While the variation could be attributed to a brief outburst of an X-ray nova
that was in its quiescence in June 2001, the X-ray nova picture is excluded
based on 20 years of observations with Einstein, ROSAT and Chandra
observatories, because the source has been brighter than $10^{39}$ erg/sec for
most of the time despite of the dramatic variations.
If the two hour period is interpreted as the orbital period, the ULX is a LMXB
system with a dwarf secondary of mass $\sim0.3 M_\odot$ filling its Roche Lobe
and accreting mass onto the primary.
%


ULX-3 has been observed with HST/WFPC2 in filters F450W (B), F555W (V) and
F814W (I).
To register ULX-3 on the optical images, we utilize a nearby X-ray source, X12
in ObsID 1622, which is on the WF3 chip and far away from the spiral arm or any
star forming regions.
Two objects are found within the $3\sigma$ error circle of X12 on the optical
images, one is a bright foreground star $1\farcs7$ away from the nominal
Chandra position, another a faint object $0\farcs4$ away from the Chandra
position with $B=24.54$ mag, $V=24.29$ mag, and $I=24.04$ mag as calculated
with HSTphot.
With the extinction corrected for the Galactic HI column density of
$1.6\times10^{20}$ cm$^{-2}$, the faint object has $M_B=-5.00$ mag, $M_V=-5.23$
mag, and $M_I=-5.44$ mag. This is likely a field globular cluster given its
isolated position and its red colors.
The X-ray spectrum of X12 shows substantial flux above 1 KeV, with
$F_X$(1--2KeV)/$F_X$(0.3--1KeV) = $0.29\pm0.15$.
This is significantly harder than X-ray spectra of stars, for which there is
essentially no emission above 1 KeV.
Thus we identify the faint object as the optical counterpart for X12, which is
presumably a LMXB system in the globular cluster, and register the ULX on the
optical images based on this identification.
The derived optical position for ULX-3, together with the X-ray and optical
positions for X12, are listed in Table 3.  The resulted error circle is of
radius $\sim0\farcs3$ (Figure 3).


ULX-3 falls on the rim of a young star cluster in the chain of star forming
knots on the spiral arm of M51. The error circle is coincident with what
appears to be a cavity between stars, with two faint objects within and a few
on the edge.
While the faint objects may not be the secondary of the ULX system which is a
$\sim0.3M_\odot$ dwarf and well below the detection limit, they, along with
other stellar objects in the immediate vicinity of ULX-3, provide an age
estimate for the system.
The stellar objects within $0\farcs5$ of the corrected X-ray position are
mostly red supergiants, as seen in the color-magnitude diagram for them and
other stellar objects from the star cluster (Figure 4).
The isochrones (Z=0.020) are overlayed on the diagram, and we found that these
stellar objects span an age range of $10^7$--$10^{7.8}$ years.
It is interesting to note that the brightest stars have left the main sequence,
and all have ages above $\sim10^7$ years, which may imply the star formation
ended 10 million years ago.


ULX-5 have been observed in the second and third Chandra observations, with an
average X-ray position R.A.=13h30m07.6s, Decl.=47d11m05.9s.
Terashima \& Wilson (2004) found that the spectra in both observations are
better fitted with an absorbed power-law model (with $n_H\sim1\times10^{21}$
cm$^{-2}$) than a multi-color disk model.
The spectrum flattened ($\Gamma=2.26$ to $\Gamma=1.86$) and the luminosity
dropped by 40\% from $3.1\times10^{39}$ erg/sec to $1.8\times10^{39}$ erg/sec
between the two observations.
If the power-law spectra are analogous to the Low Hard state in Galactic black
hole candidates and the state transition is governed by the mass accretion
rate, the ULX would be emitting at less than a few percent of the Eddington
luminosity, and a black hole of $>$ a few $\times 100 M_\odot$ is required.
If the power-law is from the Comptonization of thermal emission from the
accretion disk with a high accretion rate and the ULX is emitting at 10\% of
the Eddington level as in the case of GRO J1655-40 (Kubota et al. 2001), the
black hole may have a mass $\sim120 M_\odot$.
However, it is unclear whether the power-law spectrum is operated by the above
mechanisms, and whether the Eddington level of the ULX emission is valid for
higher mass parallels for Galactic black hole candidates. 
Thus no concrete mass constraints can be placed on the primary of the ULX
system from the X-ray spectra.


ULX-5 has been observed with HST/WFPC2 in filters F450W (B), F555W (V) and
F814W (I).
There is no other nearby X-ray sources that have optical counterparts on the
WFPC2 images, so we overlay the X-ray position directly on the optical images
as shown in Figure 5.
The ULX is on a spiral arm, and within the $0\farcs8$ error circle are seven
faint objects, the magnitudes of which are computed with HSTphot and listed in
Table 4.
The absolute $M_V$ magnitudes of these candidates for the optical counterpart
of ULX-5 have a range of -4$\sim$-5 mag, and their $B-V$ colors range from -0.1
to 0.7 mag. 
As seen from the color-magnitude diagram (Figure 6), these candidates are
evolved blue or red supergiants, and have an age range of
$10^{7.4}$--$10^{7.8}$ years.
Similar to the stellar population around ULX-3, there seems to be a lack of
bright stars younger than 10 million years in the vicinity of this ULX.


The positional accuracy of ULX-5 can be greatly improved with the help of a
nearby X-ray source, X1 in ObsID 1622, that is 40$^{\prime\prime}$ away from
ULX-5.
While this X-ray source appears to have an optical counterpart on a low spatial
resolution optical image taken with the MDM 2.4m telescope, it is unfortunately
not within the field of view of our HST observations.
Future observations with HST/ACS can be designed to cover both X1 and ULX-5,
and the relative positional accuracy can be improved to $\sim0\farcs2$. 
Given the stellar density around ULX-5 (seven objects within the $0\farcs8$
error circle), such a small error circle would be able to identify one of the
candidates as the unique optical counterpart for ULX-5, or none if the
secondary is a low mass dwarf star below detection limits.


For the stellar populations around both ULXs, interestingly there seems to be a
lack of very bright stars younger than 10 million years.
It is unlikely for this lack to be a defect caused by the photometry package or
the mis-calibration of the instruments, since such a lack is not present for the
stellar population in NGC4559, for which we used the same instrument and
processed with the same procedures.
It is difficult to diminish the lack by de-reddening the stars for the
extinction, because by making the bright stars bluer and younger, the faint
stars would be too blue to be predicted by the stellar models.
The metal abundance of the star forming regions may affect our conclusion since
stars with lower metalicity are bluer and brighter and have a shorter life.
However, the lack of young bright stars persists if we change the metalicity,
though lack of stars younger by slightly different ages.
For example, if the metalicity is lower and Z=0.001, the stellar populations
lack bright stars younger than 15 million years. If the metalicity is higher
and Z=0.100, the stellar populations lack brighter stars younger than 6 million
years.
If the lack is real, it indicates that star formation spontaneously stopped 10
million years ago for some reasons.
Alternatively, this could be a statistical defect due to small numbers of
bright stars.


\subsection{A ULX in NGC1313}

NGC1313 is a barred Sd spiral galaxy with scattered star forming regions
outside its $D_{25}$ isophote at a distance of 3.7 Mpc (Tully 1988).
Three bright X-ray point sources have been found in this galaxy, including
SN1978K and two ULXs.
Here we study one of the ULX, ULX-3 in Liu \& Bregman (2005), which is
$5^\prime$ south of the galactic nucleus and in an isolated star forming
region.

This ULX, along with SN1978K, has been observed many times with EINSTEIN,
ROSAT, and recently Chandra and XMM.
Zampieri et al. (2004) studied the long term variability of this ULX with all
available observations, and found that this ULX is highly variable, and
variability of up to a factor of 2 on a timescale of months is clearly present
which is reminiscent of the behavior observed in Galactic X-ray binaries.
The X-ray spectrum from an XMM observation in October 2000 has been studied and
it is found to be better fitted with an absorbed multi-color disk plus a
power-law model (Miller et al. 2003; Zampieri et al. 2004).
The fit gives an HI column density of $3\times10^{21}$ cm$^{-2}$ and a color
temperature of $\sim150$ eV, which if interpreted as the inner disk temperature
implies an primary black hole of $M_\bullet\sim10^3 M_\odot$.

Two Chandra observations are available in the archive, one observed in October
2002 (ObsID 2950, 20 kiloseconds), another in November 2002 (ObsID 3550, 15
kiloseconds).
The average X-ray position is R.A.=03h18m22.27s, Decl.=-66d36m03.8s with an
uncertainty of $0\farcs7$.
The optical counterpart lies in a elongated bubble nebula with a diameter of
$\sim400$ pc (Pakull \& Mirioni 2002), which shows an abrupt change in the
absolute and relative intensity of the emission lines from east to west
indicating variations in the physical conditions and/or geometry of the
emission nebula (Zampieri et al 2004).
By overlaying the X-ray position on a Bessel R band image taken with the 3.6m
telescope of ESO at La Silla in January 2002, Zampieri et al. (2004) found a
faint object with $R=21.6\pm0.2$ mag within the error circle.
Such an identification was backed up by relative astrometry using nearby
SN1978K.

We have observed the same region in the CTIO-I filter with IMACS mounted on the
Magellan/Baade telescope.
The observation has a better seeing ($\sim0\farcs6$) than the 3.6m ESO image
($\sim1^{\prime\prime}$), and is able to resolve the object B in the 3.6m ESO
image into two objects B1 and B2 (Figure 7).
The CTIO-I magnitudes were calculated with a 5-pixel aperture and calibrated
with the photometric standards in the field L98, with a photometric calibration
dispersion of 0.05 mag.
The uncertainties in the magnitudes are estimated to be 0.2 mag.
The CTIO-I magnitudes for the optical counterpart and nearby objects are listed
in Table 5 in comparison to the R magnitudes from Zampieri et al. (2004).

The optical counterpart has $M_R = -6.2$ mag and $M_I = -4.5$ mag, both within
the range of the absolute magnitudes for supergiants.
It is a very interesting blue object in that R-I$_{CTIO}$ $\approx$ -1.7 mag,
while the bluest stellar objects have R-I $\approx$ -0.32 mag (Allen, 2000).
It is even bluer after correcting for the extinction by the Galactic HI column
density of $4\times10^{20}$ cm$^{-2}$ ($A_R = 0.179$ mag, $A_I = 0.130$ mag). 
If a maximum total HI column density of $3\times10^{21}$ cm$^{-2}$ is assumed,
the correction is $A_R=1.344$ mag and $A_I=0.975$ mag.
While the difference between the CTIO-I filter and Bessel I filter may lead to
non-zero color terms, a color term of $\sim2$ mag is not expected. 
This over-blueness of the counterpart may be attributable to non-thermal
emission in the R band, e.g., extremely strong $H_\alpha$ emission line.
This R-band excess may also be caused by dramatic variability between two
observations in two years, or simply by errors in one of the measurements or
both.
We note the CTIO-I magnitude is not consistent with the R magnitude for object
A, which is a G--M supergiant as seen from its low resolution spectrum
(Zampieri et al.  2004) and should have R-I $\sim$ 0.5 mag for a G8 supergiant
or 1.1 mag for an M2 supergiant (Allen,2000).
If the R-I color is corrected by $\sim1$ mag, the optical counterpart would
have R-I $\sim$ -0.7mag, bluer than the bluest stellar objects.
Further observations with more filters including R and I are required to settle
the problem.


\subsection{A ULX in NGC628}

NGC628 is a face-on Sc spiral galaxy with considerable grand design spiral
structures at a distance of 9.7 Mpc (Tully 1988).
This galaxy was observed in June 2001 (ObsID 2057, 46 kiloseconds) and in
October 2001 (ObsID 2058, 46 kiloseconds) with the Chandra Observatory, and in
February 2002 (ObsID 0154350101, 36.9 kiloseconds) and in January 2003 (ObsID
0154350201, 24.9 kiloseconds) with the XMM-Newton Observatory, which revealed a
ULX $\sim2.^\prime5$ from the galactic nucleus, with an average X-ray position
as R.A.=01h36m51.1s, Decl.=15d45m46.8s, on a chain of star forming knots that
strays away from a main spiral arm.

This ULX showed quasi-periodic oscillations (QPO) with a quasi-period of
4,000-7,000 seconds (Liu et al. 2004).
Its QPO is  unique behavior among ULXs and Galactic X-ray binaries due to the
combination of its burst-like peaks and deep troughs, its long quasi-periods,
its high variation amplitudes of $>90$\%, and its substantial variability
between observations. 
The X-ray spectra in both Chandra observations are fitted by an absorbed
accretion disk plus a power-law component, suggesting the ULX was in in a
spectral state analogous to the Low Hard state or the Very High state of
Galactic black hole X-ray binaries.  
A black hole mass of $\sim2$--$20\times10^3 M_\odot$ is estimated from the
$f_b$--$M_\bullet$ scaling relation found in the Galactic X-ray binaries and
active galactic nuclei (Liu et al. 2004).

The ULX was observed in Bessel B, Bessel V, and CTIO-I filters with IMACS in
search of its optical counterpart.
A few X-ray sources have optical counterparts on the IMACS images, showing that
the Chandra positions are accurate to better than $0\farcs6$ without systematic
offsets.
A direct overlay of the X-ray position on the optical images (Figure 9) reveals the ULX is
surrounded by a blue shell structure, possibly a bubble nebula of $\sim300$pc
as seen in the case of the ULX in NGC1313.
An extended fuzzy object (C1) is marginally discernible at the nominal Chandra
position, and could be a supergiant star that is the optical counterpart of the
ULX. The counterpart could also be another brighter object (C2) which is
$\sim0\farcs4$ away from the nominal Chandra position.

The magnitudes for C1, C2, and three other objects within the $3\sigma$ error
circle of the X-ray position of the ULX are listed in Table 6.
The uncertainties are estimated to be as large as 0.4 mag owing to the
crowdedness of the objects and the diffuse background.
After the extinction is corrected for the Galactic HI column density of
$\sim5\times10^{20}$ cm$^{-2}$ ($A_B$=0.352 mag, $A_V$=0.265 mag, $A_I$=0.163
mag), the $M_V$ magnitudes for these objects range from -7.3 mag to -9 mag,
which is $\sim0.7$--$3.4\times10^5$ times brighter than the Sun.
The $B-V$ colors range from 0.03 mag to 0.23 mag, the same range for A0 -- A8
stars. 
Given their extendedness, their absolute magnitudes and their colors, these
objects are likely young star associations/clusters in the star forming region.
However, any definite conclusions are premature given the large uncertainties
in the magnitudes and colors, and the large seeing ($0\farcs6\sim$30 pc) for
the ground-based observations.
Further optical observations with HST are required to clarify the nature of
these objects.


\section{DISCUSSION}


An essential step in understanding the nature of the enigmatic ULXs is to study
them in other wavelengths, especially in optical.
Along this line of thought, we have studied a sample of ULXs in optical with
HST and the Magellan/Baade 6.4m telescope.
These include the ULX in NGC3031, which is identified with an O8V star (Liu et
al. 2002), and the ULX in NGC5204, which is identified with a B0Ib supergiant
with considerable FUV emission from the accretion disk (Liu et al. 2004).
Optical studies of five ULXs in four galaxies are reported in this paper. 
The ULX in NGC4559 is identified with an OB association in an isolated star
forming complex on the outer edge of NGC4559.  
The ULX-3 in NGC5194 is located on the rim of a young star cluster on a spiral
arm, and its probable secondary of $\sim0.3 M_\odot$ as inferred from its two
hour period is below the detection limit. 
The ULX-5 in NGC5194 is located on a spiral arm, and identified with a few blue
and red supergiants. 
The ULX in NGC1313 is identified to an object with R-I color bluer than the
bluest stellar objects, indicative of non-thermal emission in the R band, or
dramatic variability between two observations in two years, or errors in
measurements. 
The ULX in NGC628 is located in a bubble nebula of $\sim300$ pc on a chain of
star forming knots, and identified with a few young stellar
associations/clusters.
%


Optical studies have revealed a strong tendency for ULXs to occur in very young
star forming regions, which suggests that ULXs are associated with the
production and rapid evolution of short-lived massive stars.
Pakull \& Mirioni (2002) surveyed fifteen ULXs in eleven nearby galaxies with
ground based telescopes, and found HII regions, (bubble) nebulae, and X-ray
ionized nebulae (XIN) within the error circles of the ROSAT HRI positions for
thirteen ULXs.
Our optical studies of the ULXs show that their stellar environments are very
young, for example, younger than $10^{7.6}$ years for the ULX in NGC3031,
younger than $10^{7.6}$ years for the ULX in NGC4559, and younger than
$10^{7.8}$ years for the two ULXs in NGC5194.
These findings are consistent with the model that ULXs are HMXB systems
undergoing thermal timescale mass transfer through Roche lobe overflow, which
is a bright, short-lived, but common phase that must follow the familiar
wind-fed phase of HMXBs, as is probably seen in SS433 (King et al. 2002).

The close connection between ULXs and HMXBs is dramatically demonstrated by the
recent Chandra observation of the Cartwheel galaxy (Gao et al. 2003).
The Chandra observation reveals more than 20 ULXs with $L_{0.5-10 keV} \ge
3\times10^{39}$ erg/sec, most of which are confined in an expanding ring, the
main site of recent massive star formation that started $3\times10^8$ years
ago.
The lack of radial spread of the ULXs suggests these ULXs must have ages
$\le10^7$ years given the expansion velocity of the ring (Gao et al. 2003).
King (2004) further points out that the initial secondary mass must be $\ge15
M_\odot$, and the total number of HMXBs in the past $3\times10^8$ years is
$\ge3000$ given the lifetime of these HMXBs ($\le10^7$ years), the beaming
factor ($b\le0.1$), and the duty cycle ($d\sim1$).
If ULXs are IMBHs, the systems are likely to be transient with unstable
accretion disks that probably have duty cycles $d\le10^{-2}$. 
It would be problematic to explain the observation with the IMBH model, since
it is unlikely for this galaxy to produce the required $\ge3\times10^4$ IMBHs
to account for the observed ULXs.

Direct support for the HMXB scenario comes from the optical identifications of
ULXs with massive stars as secondaries of the ULX systems.
Only four ULXs have been identified with unique counterparts so far.
A ULX in NGC3031 is identified with an O8 V star within the $0\farcs2$ error
circle (Liu, Bregman, \& Seitzer 2002). 
A ULX in NGC5204 is identified with a B0 Ib supergiant within the $0\farcs2$
error circle (Liu et al. 2004).
A ULX in Holmberg II is identified with a point-like object within the
$0\farcs6$ error circle, for which the absolute magnitude and B-V color are
consistent with a range of spectral types from O4V to B3 Ib (Kaaret, Ward \&
Zezas 2004).
A ULX in NGC1313 is identified with a point-like object within the $1\farcs4$
error circle, for which the absolute R magnitude is consistent with an early O
star or an OB supergiant (Zamperi et al. 2004).
Note that the optical identifications are usually based on positional
coincidence, and may be nearby stars not related to the ULXs, especially in
dense stellar fields with large error circles.
The identification, therefore, is not secure until verified by other
independent means.
The identification for the ULX in NGC5204 is so far the only one verified by
the spectroscopic observation with HST/STIS, which reveals, in addition to
typical B0 Ib spectral features, an NV emission line suggestive of the
existence of a hot accretion disk (Liu et al.  2004).


Once a ULX is identified with a stellar object as the secondary, follow-up
spectroscopic observations of the secondary can determine its physical
properties, its radial velocity curve and ultimately the mass of the primary.
With medium/high resolution spectroscopy, one can identify spectral features
and improve the spectral type estimated from the wide-band magnitudes and
colors.
In addition, possible spectral features from the accretion disk, e.g., the high
ionization emission line of NV as in the case of NGC5204, can be used to secure
the identification, or to elect the counterpart when more than one object are
within the error circle as in the case of NGC4559.
Once the mass ($M_1$) and size ($R_1$) of the secondary is determined
empirically from its spectral type, the orbital properties will depend solely
on the mass of the primary black hole ($M_2$) if the secondary is presumably
filling its Roche lobe, for which the radius is $R_{cr} = {0.49 A \over 0.6 +
q^{-2/3} ln(1+q^{1/3})}$ (Eggleton 1983).
Here $A$ is the orbital separation, and $q = M_1/M_2$ is the mass ratio.
With known $M_1$ and $R_{cr} = R_1$, one finds that the separation $A$, thus
the orbital period $P$, depends only on the primary mass $M_2$ given the
Kepler's Law $P^2/A^3 = 4\pi^2 G(M_1+M_2)$.
The rotational velocity of the secondary also depends on the primary, with
$v_1^2 = GM_2^2(M_1+M_2)/A$.

The expected periods and rotational velocities are calculated as a function of
the primary mass for a secondary of B0 Ib supergiant with $M_1 = 25 M_\odot$
and $R_1 = 30 R_\odot$ as an example (Figure 10).
For the primary mass increases from 3 $M_\odot$ to $10^3 M_\odot$, the orbital
period increases from $\sim$200 hours to $\sim$300 hours at $100 M_\odot$ then
decreases to $\sim$280 hours, while the rotational velocity of the secondary
increases monotonically from 30 km/sec to $10^3$ km/sec.
Doppler shifts of such rotational velocities are easy to detect as they in a
edge-on viewing geometry lead to a line shift of 0.45($\lambda/4500$\AA) for a
3 $M_\odot$ black hole, 1.5 ($\lambda/4500$\AA) for a 100 $M_\odot$ black hole,
and 15 ($\lambda/4500$\AA) for a $10^3 M_\odot$ black hole.
By taking a series of spectroscopic observations and constructing a radial
velocity curve of the secondary, we would be able to determine the mass of the
primary black hole, or the lower limit of the mass if the ULX system is not
viewed edge-on.

Optical studies have also revealed a few ULXs with low mass secondaries.
These include a ULX in NGC4565 (Wu et al. 2001) and and two ULXs in the
elliptical galaxy NGC1399 (Angelini et al. 2001) identified with globular
clusters.
The secondaries in these ULXs must be low mass stars owing to lack of massive
stars in globular clusters.
Another example is the ULX-3 in NGC5194, for which the secondary mass is
estimated to be $\sim0.3M_\odot$ if the observed two hour period is its orbital
period, and if the secondary is overflowing the Roche lobe.
%
The estimated secondary mass will be lower if the secondary is not filling its
Roche lobe.
Note that it is not necessary for the secondary to fill the Roche lobe to get
high luminosities. 
King (2002) suggests that high inflow rates can be obtained in the outburst
phase of soft X-ray transients (e.g., GRS 1915+105), which exists for most
LMXBs, and inevitable for any LMXB with orbital period $\ge2$ days.
For example, the inflow rate will exceed $\sim10^{-7} M_\odot/yr$ for a
$10M_\odot$ black hole if the orbital period $\ge1$ day, and lead to a total
luminosity of $\ge10^{39}$ erg/sec.
The outburst phase is long and may span decades, but the quiescent intervals
are much longer, and the duty cycle is $\ll1$, which decreases with increasing
orbital periods (King 2002).

\acknowledgements

We would like to thank Renato Dupke, Eric Miller for helpful discussions.  We
gratefully acknowledge support for this work from NASA under grants
HST-GO-09073.



\begin{deluxetable}{lllll|cccll}
\tablecaption{The optical position for the ULX in NGC4559}
\tabletypesize{\tiny}
\tablehead{
\colhead{} & \multicolumn{4}{c}{ACIS-S3} \vline& \multicolumn{5}{c}{WFPC2} \\
\colhead{Object} & \colhead{RA} & \colhead{DEC} & \colhead{RA\_ERR} &
\colhead{DEC\_ERR} \vline& \colhead{Chip} & \colhead{X} & \colhead{Y} &
\colhead{RA} & \colhead{DEC} \\
\colhead{} & \colhead{} & \colhead{} & \colhead{(arcsec)} &
\colhead{(arcsec)} \vline& \colhead{} & \colhead{(pix)} & \colhead{(pix)} &
\colhead{} & \colhead{}
}

\startdata
X4      &12:35:48.56 & 27:55:31.5  & 0.07 & 0.17 & 2 & 584.0  & 237.0  &12:35:48.64 & 27:55:32.0  \\
ULX     &12:35:51.71 & 27:56:04.1  & 0.01 & 0.01 & 1 & 445.7  & 430.7  &12:35:51.79 & 27:56:04.6  \\
\hline
X5      &12:35:47.67 & 27:56:57.2  & 0.12 & 0.14 & 3 & 584.0  & 237.0  &12:35:47.76 & 27:56:56.4  \\
ULX     &12:35:51.70 & 27:56:04.9  & 0.03 & 0.03 & 1 & 445.7  & 430.7  &12:35:51.79 & 27:56:04.1  \\

\enddata
\end{deluxetable}


\begin{deluxetable}{llllllllll}
\tablecaption{Optical counterpart Candidates for NGC4559-ULX}
\tabletypesize{\tiny}
\tablehead{
 \colhead{}  & \multicolumn{3}{c}{app. mag.} & \multicolumn{3}{c}{5.8 Mpc} & \multicolumn{3}{c}{9.7 Mpc} \\
 \colhead{Candidate} & \colhead{$m_B$} & \colhead{$m_V$} & \colhead{$m_I$}  & \colhead{$M_B$} & \colhead{$M_V$} & \colhead{$M_I$}  & \colhead{$M_B$} & \colhead{$M_V$} & \colhead{$M_I$} \\
}

\startdata

C-1 &24.82$\pm$0.083&24.56$\pm$0.048&24.08$\pm$0.076&-3.894&-4.177&-4.69&-5.004&-5.287&-5.8 \\
C-2 &25.18$\pm$0.117&24.24$\pm$0.041&23.1$\pm$0.031&-3.529&-4.499&-5.667&-4.639&-5.609&-6.777 \\
C-3 &24.46$\pm$0.137&23.91$\pm$0.078&22.95$\pm$0.057&-4.258&-4.831&-5.817&-5.368&-5.941&-6.927 \\
C-4 &22.79$\pm$0.022&23$\pm$0.016&23.2$\pm$0.04&-5.927&-5.738&-5.575&-7.037&-6.848&-6.685 \\

\enddata
\end{deluxetable}


\begin{deluxetable}{lllll|cccll}
\tablecaption{The optical position for ULX-3 in NGC 5194}
\tabletypesize{\tiny}
\tablehead{
\colhead{} & \multicolumn{4}{c}{ACIS-S3} \vline& \multicolumn{5}{c}{WFPC2} \\
\colhead{Object} & \colhead{RA} & \colhead{DEC} & \colhead{RA\_ERR} &
\colhead{DEC\_ERR} \vline& \colhead{Chip} & \colhead{X} & \colhead{Y} &
\colhead{RA} & \colhead{DEC} \\
\colhead{} & \colhead{} & \colhead{} & \colhead{(arcsec)} &
\colhead{(arcsec)} \vline& \colhead{} & \colhead{(pix)} & \colhead{(pix)} &
\colhead{} & \colhead{}
}

\startdata
X12     &13:29:56.25 & 47:14:51.5  & 0.09 & 0.08 & 3 & 304.0  & 604.0  &13:29:56.21 & 47:14:51.5  \\
ULX     &13:30:01.01 & 47:13:44.0  & 0.15 & 0.12 & 1 & 392.5  & 400.1  &13:30:00.97 & 47:13:44.0  \\

\enddata
\end{deluxetable}


\begin{deluxetable}{llll}
\tabletypesize{\small}
\tablecaption{Optical Counterpart candidates for ULX-5 in NGC 5194}
\tablehead{
 \colhead{candidate} & \colhead{$m_B$} & \colhead{$m_V$} & \colhead{$m_I$} \\
 \colhead{}  & \colhead{mag}  & \colhead{mag}  & \colhead{mag} \\
}

\startdata

c1&25.788$\pm$0.229&25.562$\pm$0.093&25.079$\pm$0.192 \\
c2&25.751$\pm$0.163&\nodata&\nodata \\
c3&\nodata&26.052$\pm$0.222&24.946$\pm$0.174 \\
c4&24.409$\pm$0.057&24.493$\pm$0.041&24.327$\pm$0.085 \\
c5&25.803$\pm$0.242&25.142$\pm$0.128&23.033$\pm$0.038 \\
c6&24.875$\pm$0.1&24.26$\pm$0.036&23.69$\pm$0.05 \\
c7&25.357$\pm$0.138&25.376$\pm$0.086&\nodata \\

\enddata
\end{deluxetable}


\begin{deluxetable}{lllll}
\tabletypesize{\small}
\tablecaption{Optical counterpart to NGC1313-ULX and nearby objects}
\tablehead{
 \colhead{object} & \colhead{$I_{CTIO}$} & \colhead{$m_R$} & \colhead{$M_I$} & \colhead{$M_R$} \\
 \colhead{}  & \colhead{mag}  & \colhead{mag}  & \colhead{mag}  & \colhead{mag} \\
}

\startdata

A&19.8&19.8&-8.04&-8.04  \\
B1&22.8 & &-5.04  \\
B2&22.9 & &-4.94  \\
B1+B2&22.1&20.7&-5.74&-7.14  \\
C(ULX)&23.3&21.6&-4.54&-6.24  \\
D&17.1&17.8&-10.74&-10.04  \\

\enddata
\end{deluxetable}


\begin{deluxetable}{lllllll}
\tabletypesize{\small}
\tablecaption{Optical Counterpart candidates for the ULX in NGC628}
\tablehead{
 \colhead{Candidate} & \colhead{$m_B$} & \colhead{$m_V$} & \colhead{$I_{CTIO}$}  & \colhead{$M_B$} & \colhead{$M_V$} & \colhead{$M_I$} \\
 \colhead{}  & \colhead{mag}  & \colhead{mag}  & \colhead{mag}  & \colhead{mag}  & \colhead{mag}  & \colhead{mag} \\
}

\startdata

C1&22.91&22.79&22.04&-7.02&-7.14&-7.89  \\
C2&21.87&21.71&21.47&-8.06&-8.22&-8.46  \\
C3&22.65&22.43&21.89&-7.28&-7.496&-8.04  \\
C4&22.53&22.22&21.82&-7.401&-7.712&-8.11  \\
C5&21.39&21.28&20.98&-8.541&-8.652&-8.95  \\

\enddata
\end{deluxetable}


\begin{figure}
\plottwo{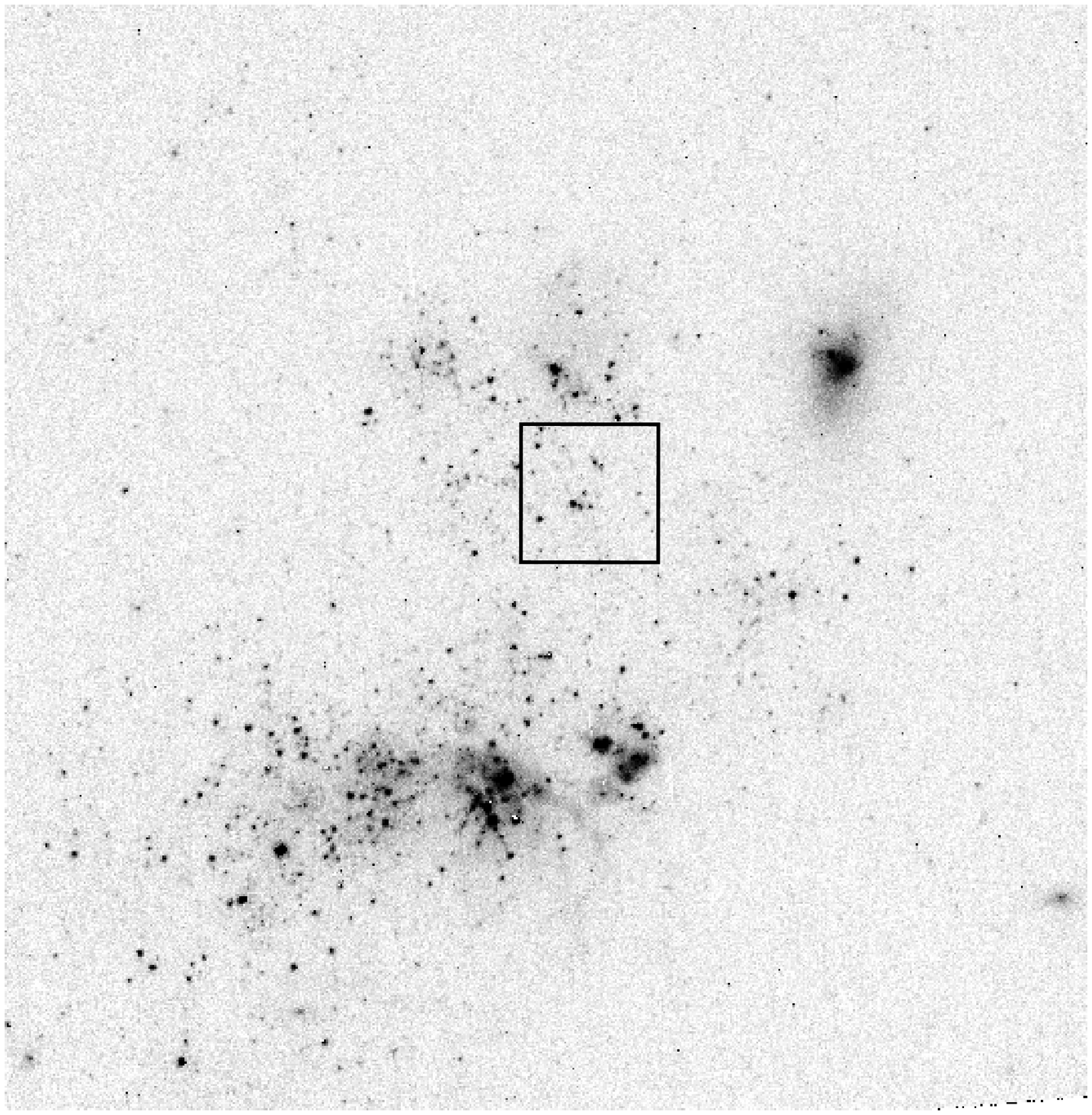}{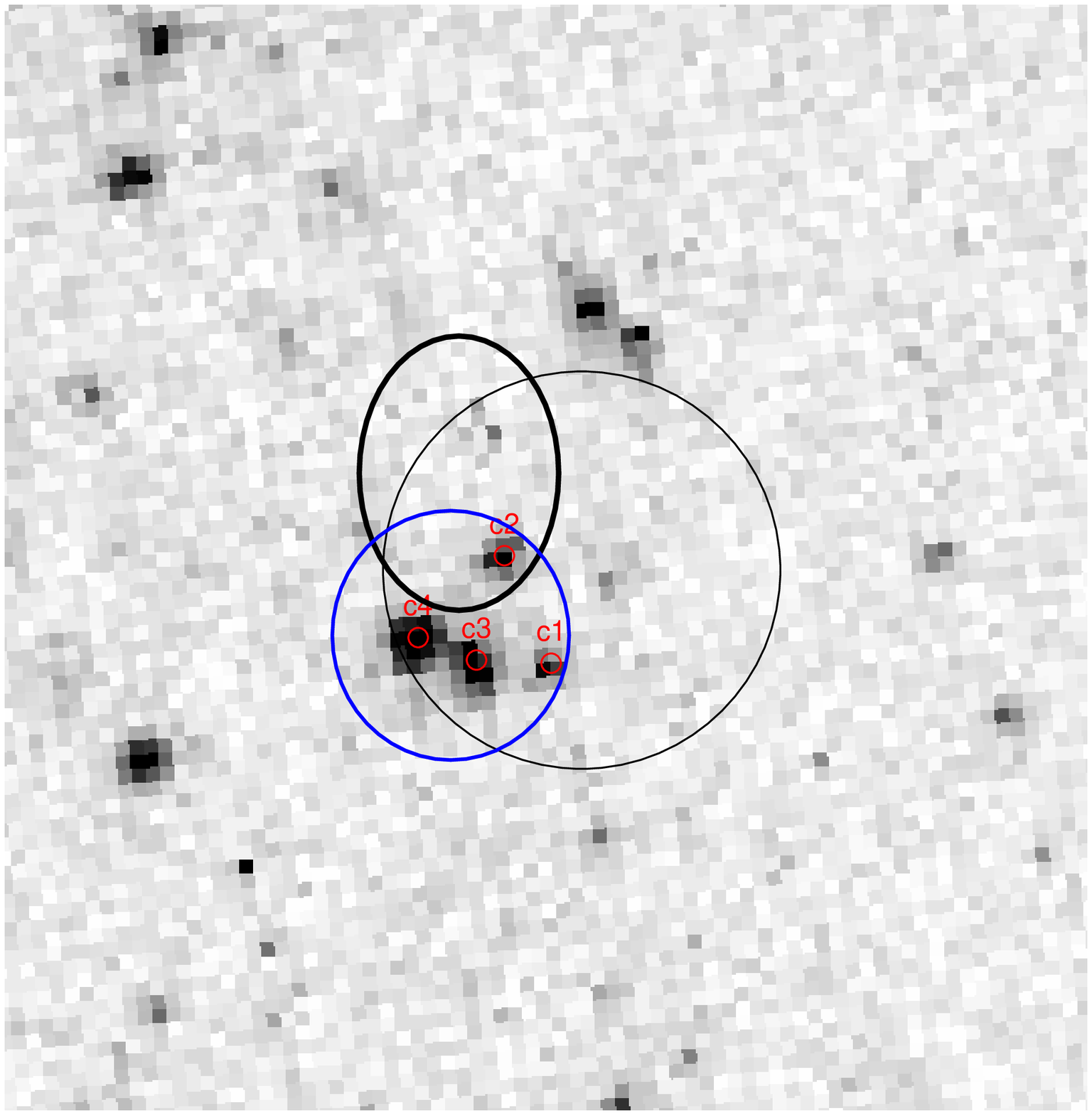}
\caption{The optical identification of the ULX in NGC4559.  The ULX, enclosed
by a box region of 80 PC pixels $\times$ 80 PC pixels in the left panel, lies
between two young star clusters on a F555W PC image. A zoom-in of the boxed
region is shown in the right panel. The $0\farcs6$ thin error circle (thin) was
adopted by Cropper et al., the thick round error circle was based on the
identification of X5 as a point source, and the error ellipse was based on the
identification of X4 as an extended source. North is up for all images in this
paper. }

\end{figure}

\begin{figure}
\plottwo{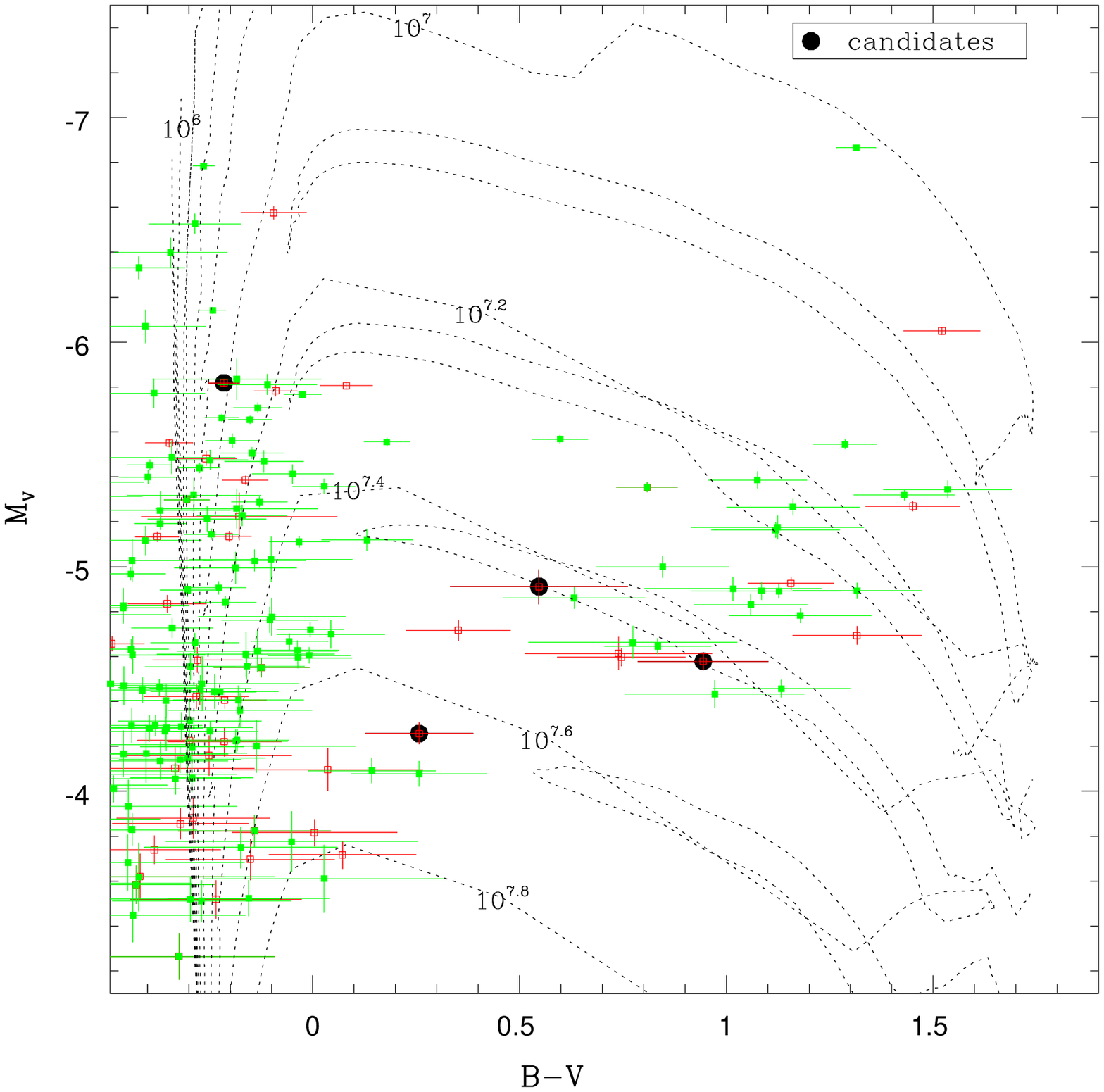}{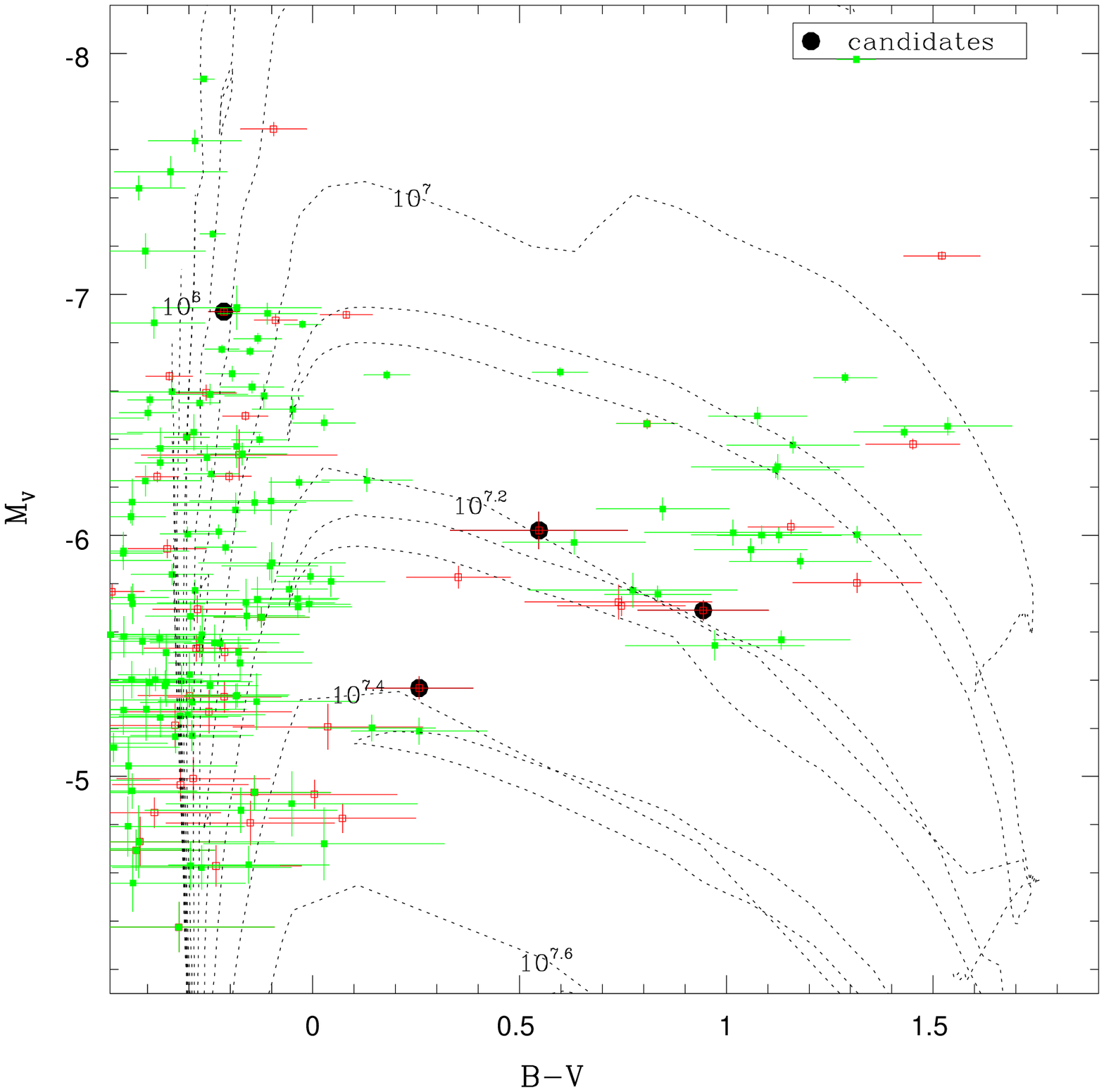}
\caption{ The color-magnitude diagrams for the four optical counterpart
candidates for the ULX in NGC4559 and the nearby stars. The distance to NGC4559
is taken as 5.8 Mpc for the left panel, and 9.7 Mpc for the right panel. In
both diagrams the isochrones from Lejeune \& Schaerer 2001 with Z=0.020 are
overplotted for comparison.    }

\end{figure}
 
\begin{figure}
\plottwo{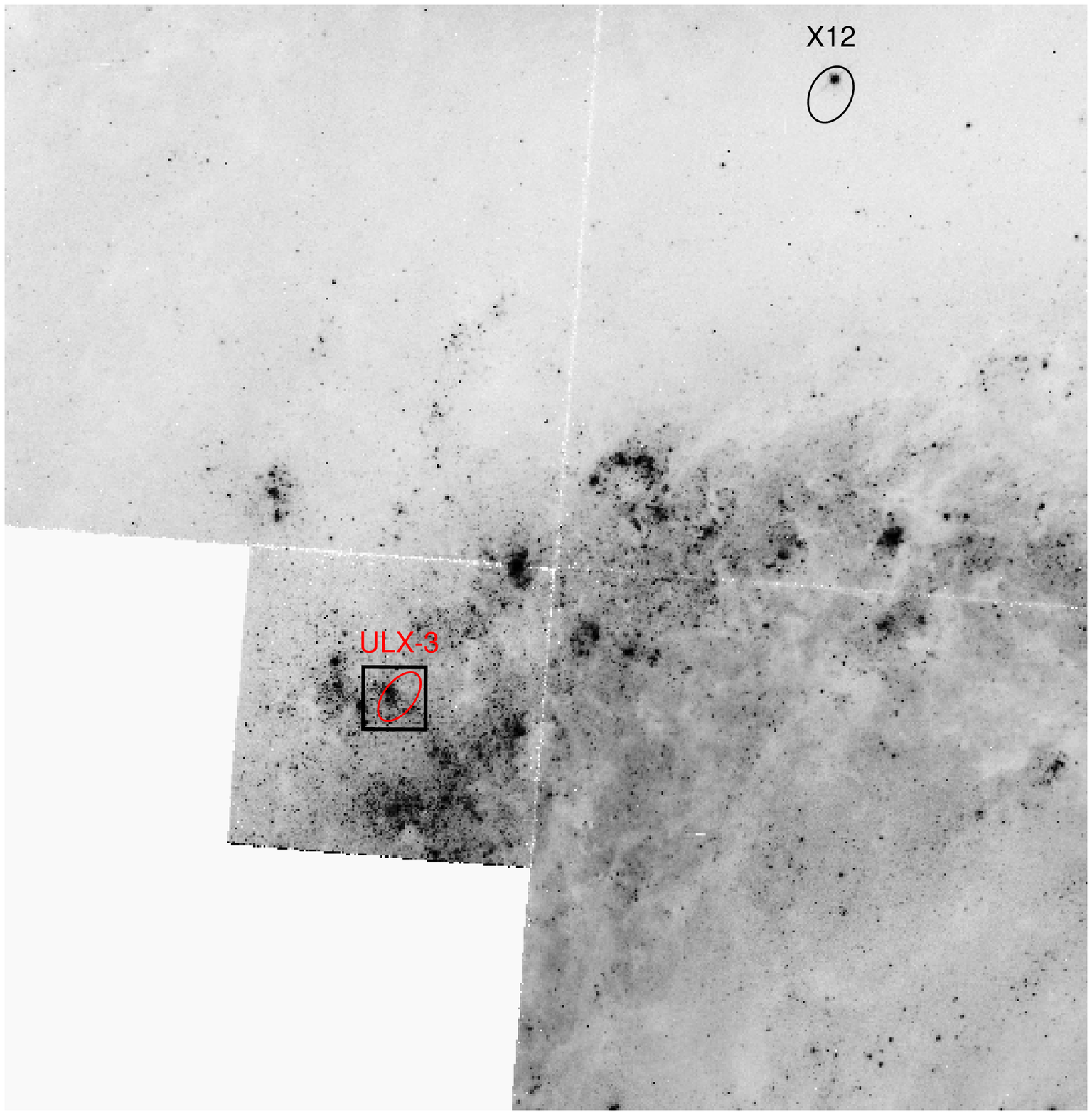}{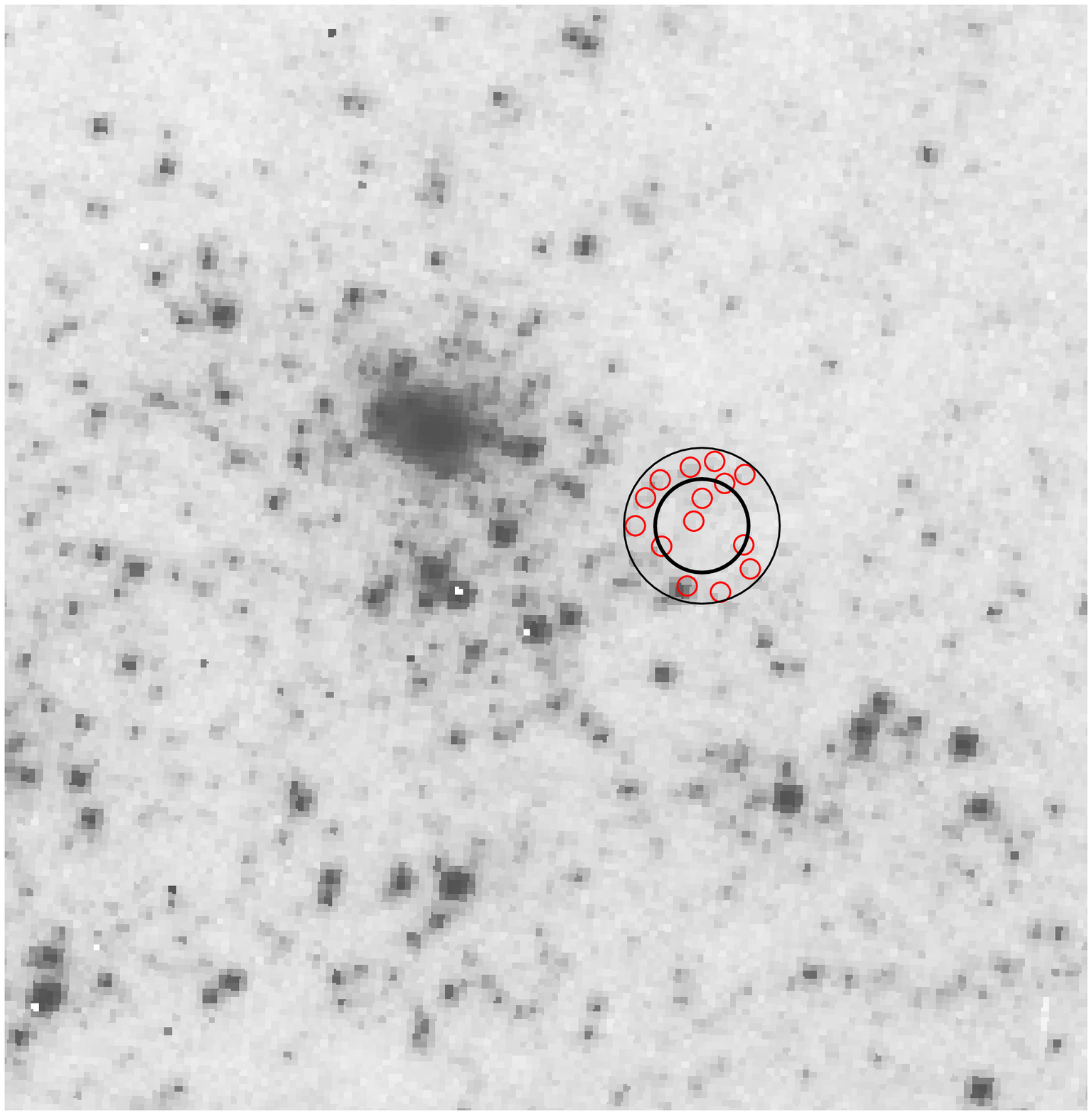}
\caption{The optical identification of the ULX-3 in NGC5194.  In the left panel,
the $3\sigma$ error ellipses of the X-ray positions of ULX-3 and X12 are
overlayed on a mosaic WFPC2 image.  The ULX position was adjusted by
identifying X12 to a possible globular cluster as described in the text. A
zoom-in of the boxed region (80 PC pixels $\times$ 80 PC pixels) is shown in the
right panel. The thick circle around the adjusted ULX position has a radius of
$0\farcs3$, and the thin circle has $0\farcs5$.  }

\end{figure}

\begin{figure}
\plotone{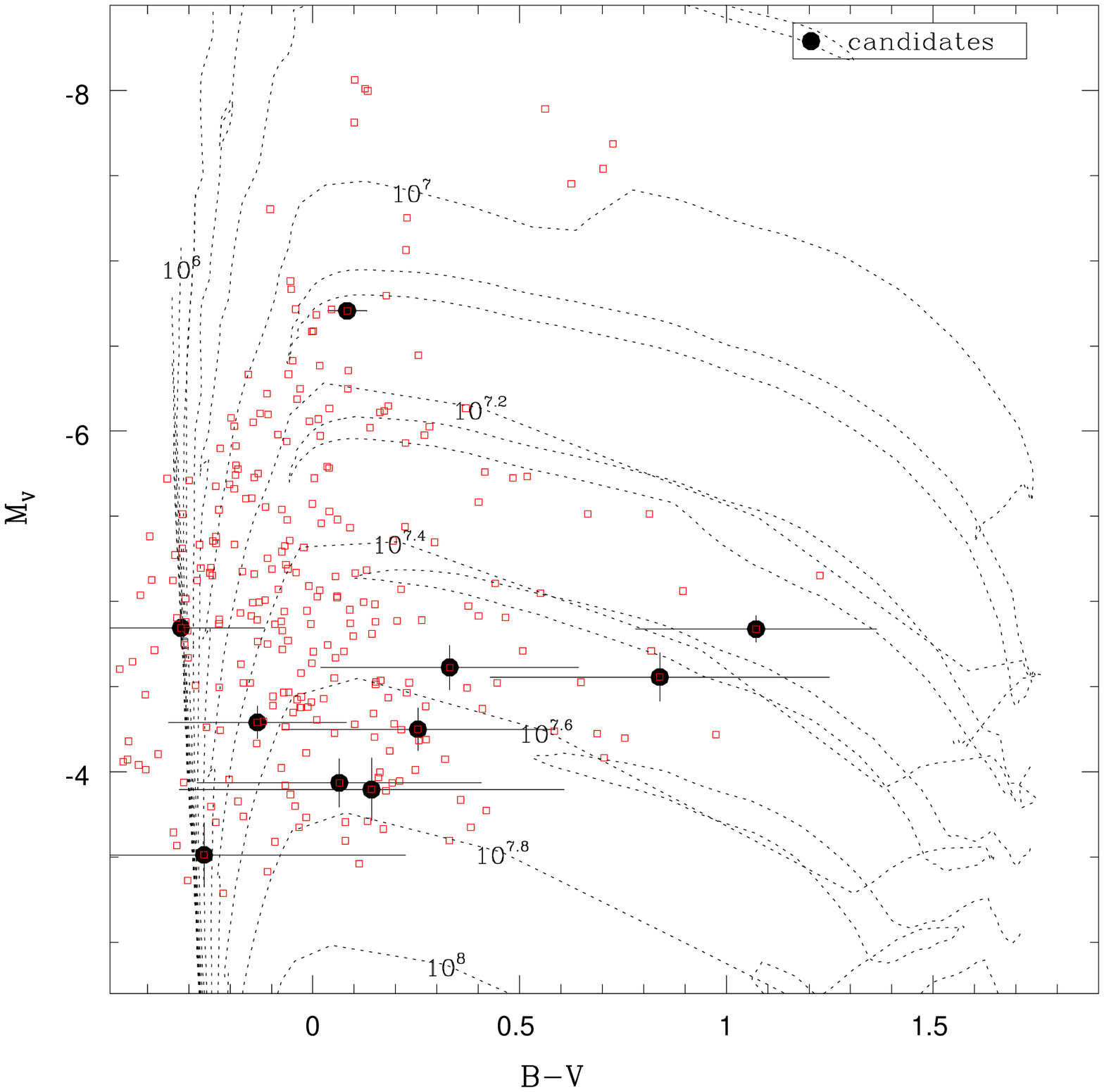}
\caption{The color-magnitude diagram for the nearby stars around the ULX-3 in
NGC5194. The filled circles are those within $0\farcs5$ of the adjusted ULX
position. The isochrones from Lejeune \& Schaerer 2001 with Z=0.020 are
overplotted for comparison.  }

\end{figure}

\begin{figure}
\plottwo{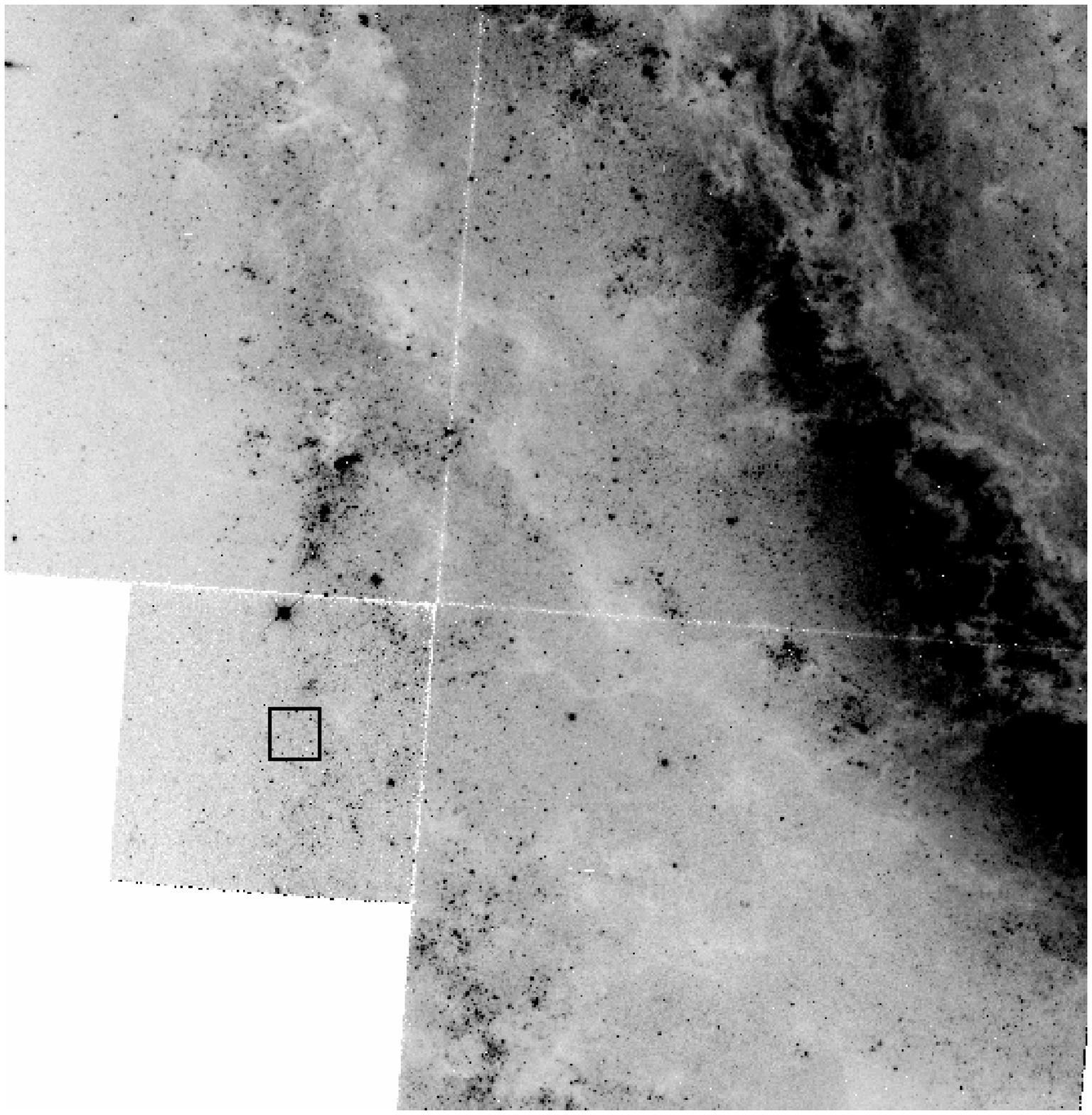}{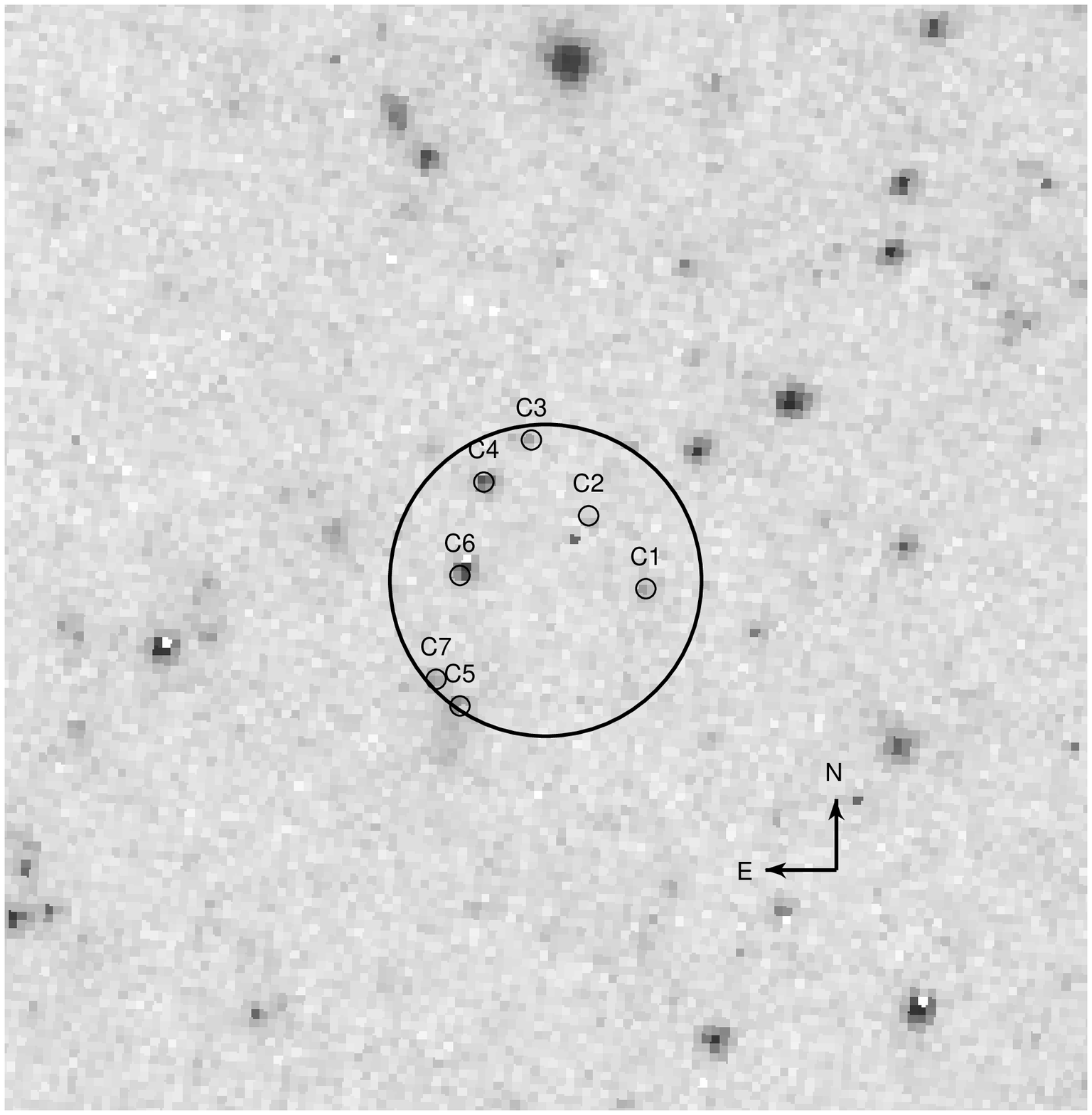}
\caption{The optical identification of the ULX-5 in NGC5194. The ULX-5, enclosed
in a boxed region of 80 pixels $\times$ 80 pixels, is on the outer edge of a
spiral arm as seen from the mosaic WFPC2 image in the left panel. A zoom-in of
the boxed region is shown in the right panel. Seven point sources are enclosed
in the $0\farcs8$ error circle around the nominal X-ray position. }

\end{figure}
 
\begin{figure}
\plotone{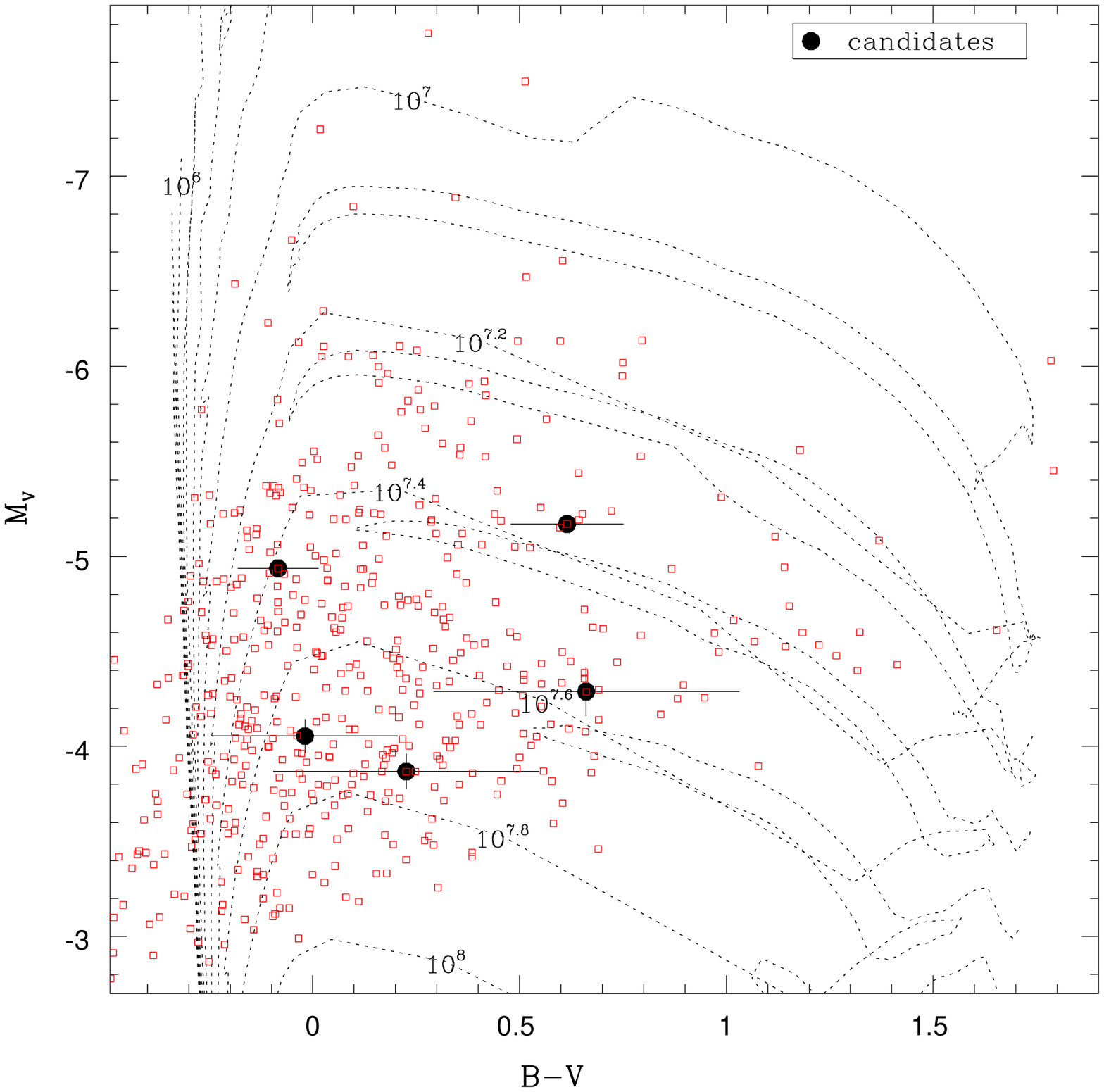}
\caption{The color-magnitude diagram for the nearby stars around the ULX-5 in
NGC5194. The filled circles are those within the $0\farcs8$ error circle of the
nominal ULX position. The isochrones from Lejeune \& Schaerer 2001 with Z=0.020
are overplotted for comparison.   }

\end{figure}
 
\begin{figure}
\plotone{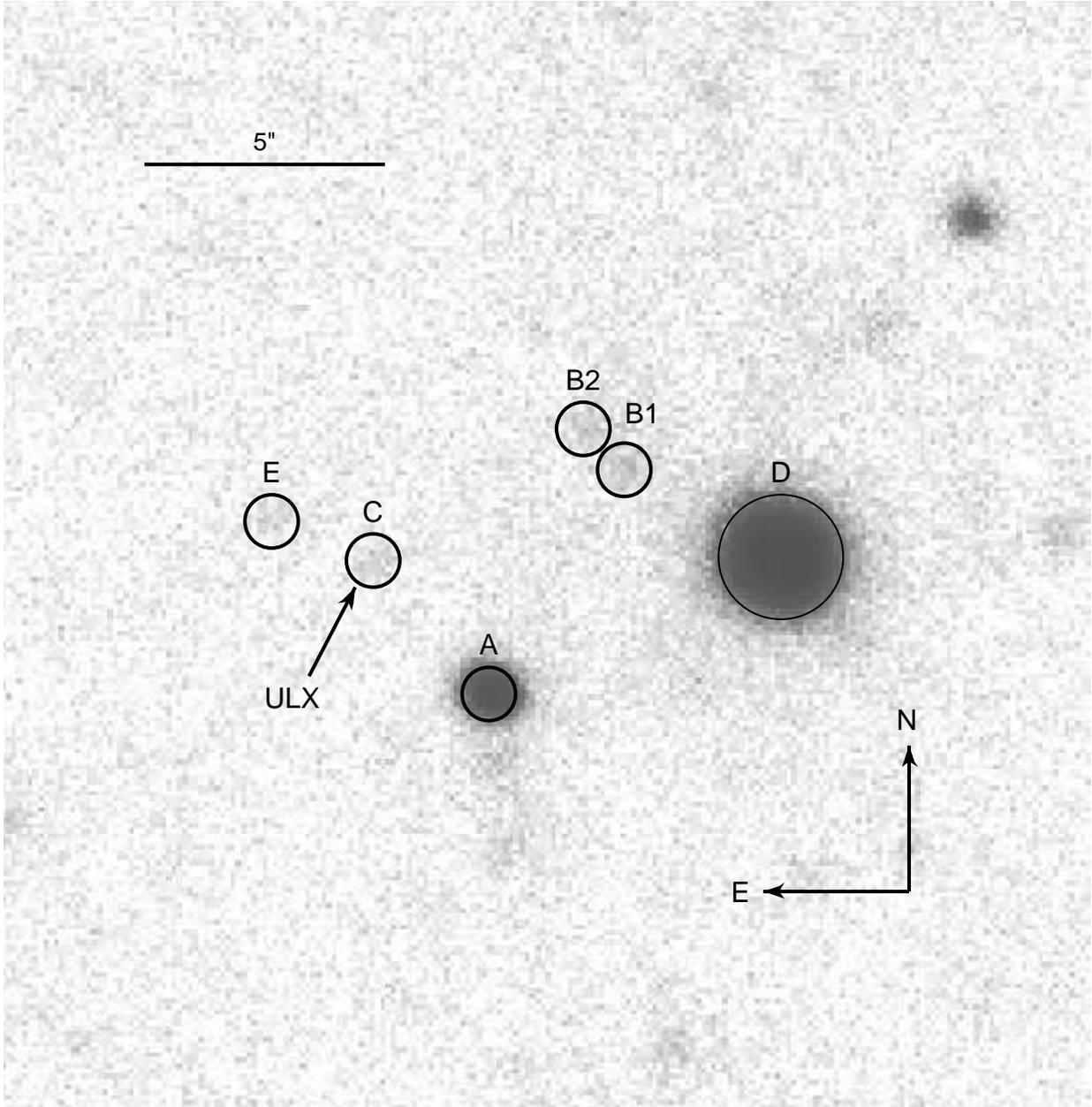}
\caption{The optical counterpart for the ULX in NGC1313 on a CTIO-I image taken
with the 6.4 meter Magellan/Baade telescope. The object names follow those in
Zampieri et al. (2004), except that the object B on their R image taken with
the 3.6 meter ESO telescope is resolved into two objects.  }

\end{figure}

\begin{figure}
\plotone{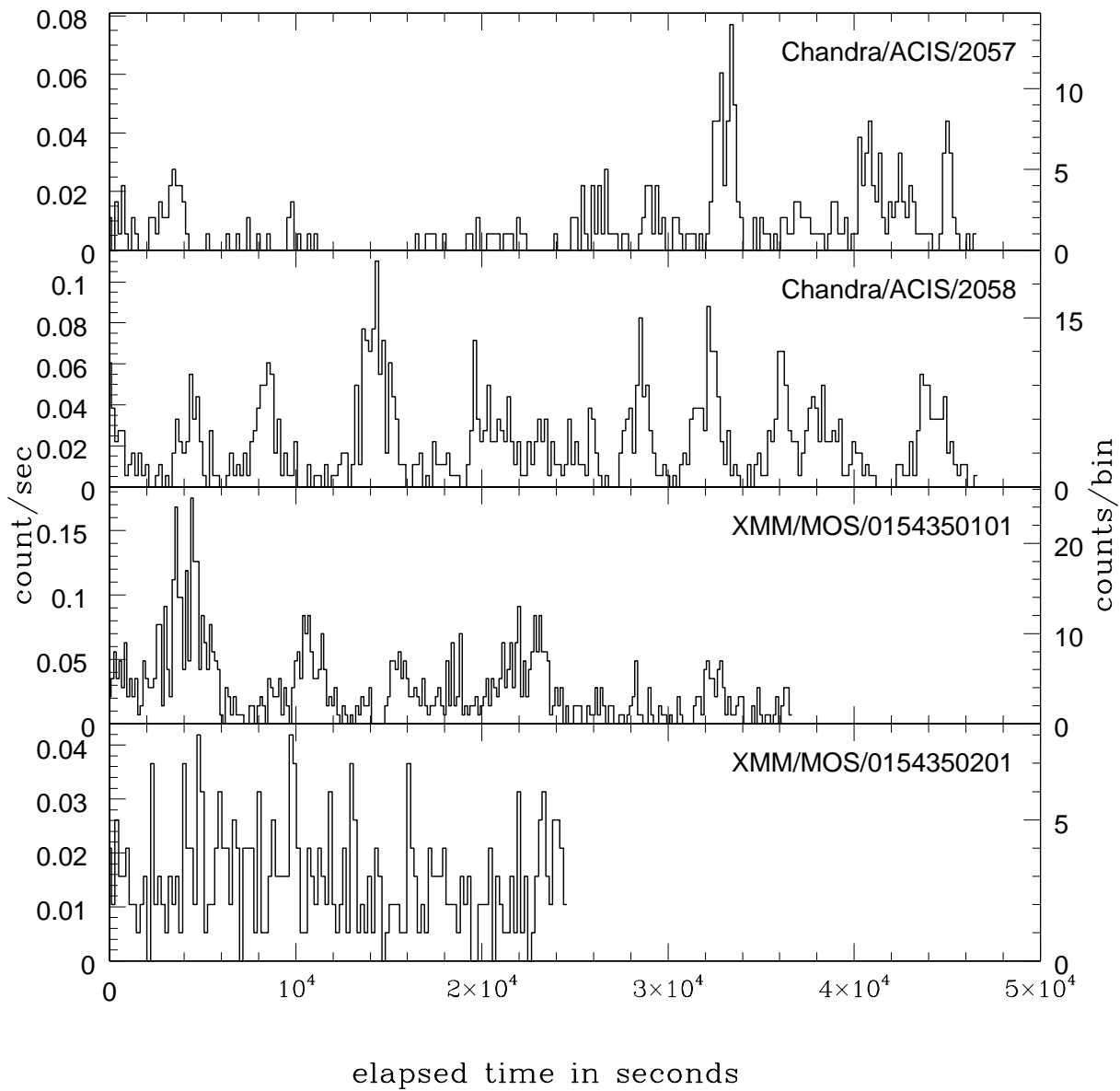}
\caption{ The light curves for the ULX in NGC628 in four observations. The
quasi-periodic oscillations are apparent from the light curves, and unique for
its burst-like peaks and deep troughs, its long quasi-period of $\sim$ 4--7
kiloseconds, its  high variation amplitudes, and its substantial variability
between observations. Overplotted for comparison are sinusoidal curves with the
periods of 4000/4000/7000/6000 seconds respectively. For details see Liu et al.
(2004).  }

\end{figure}

\begin{figure}
\plottwo{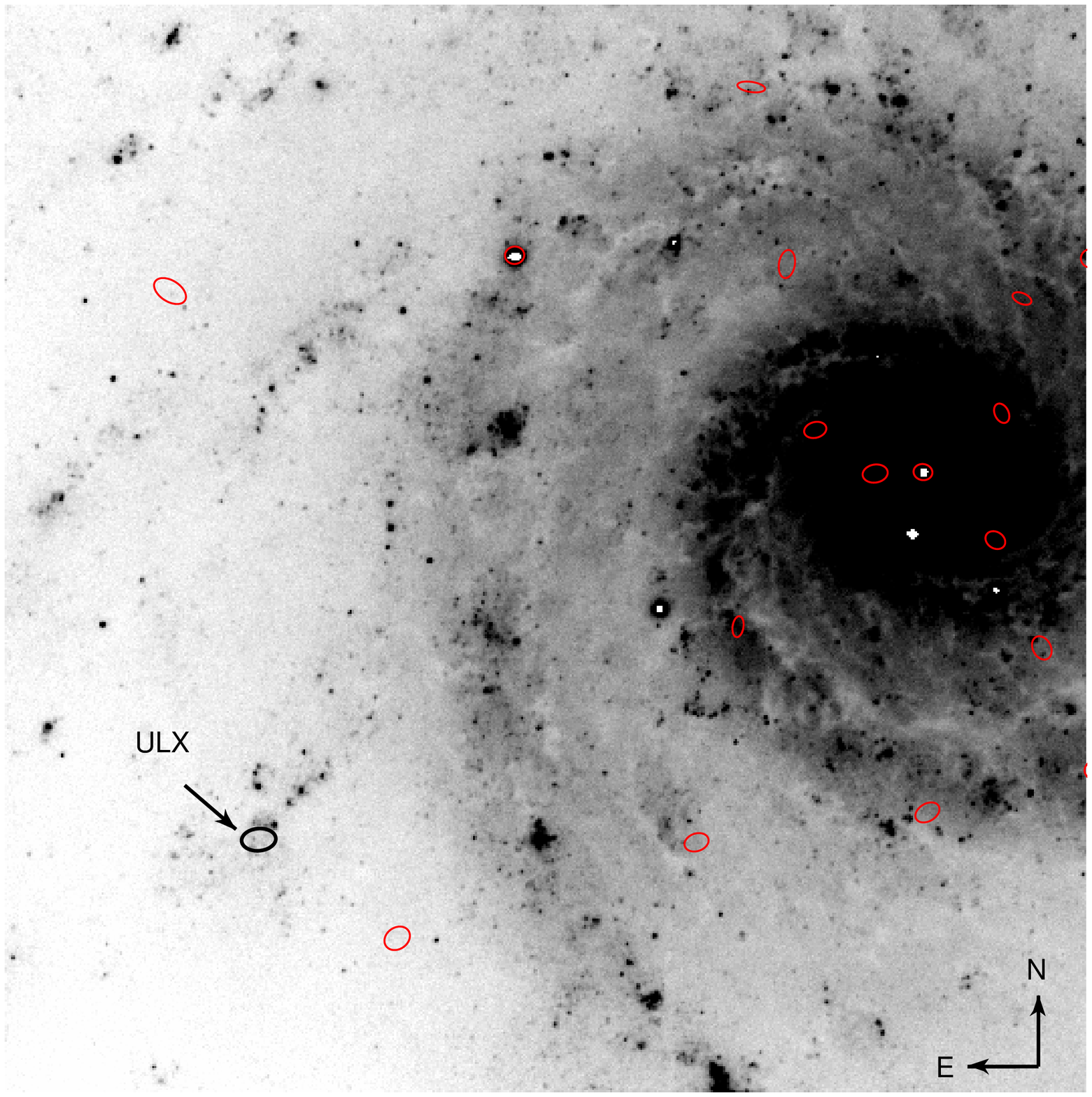}{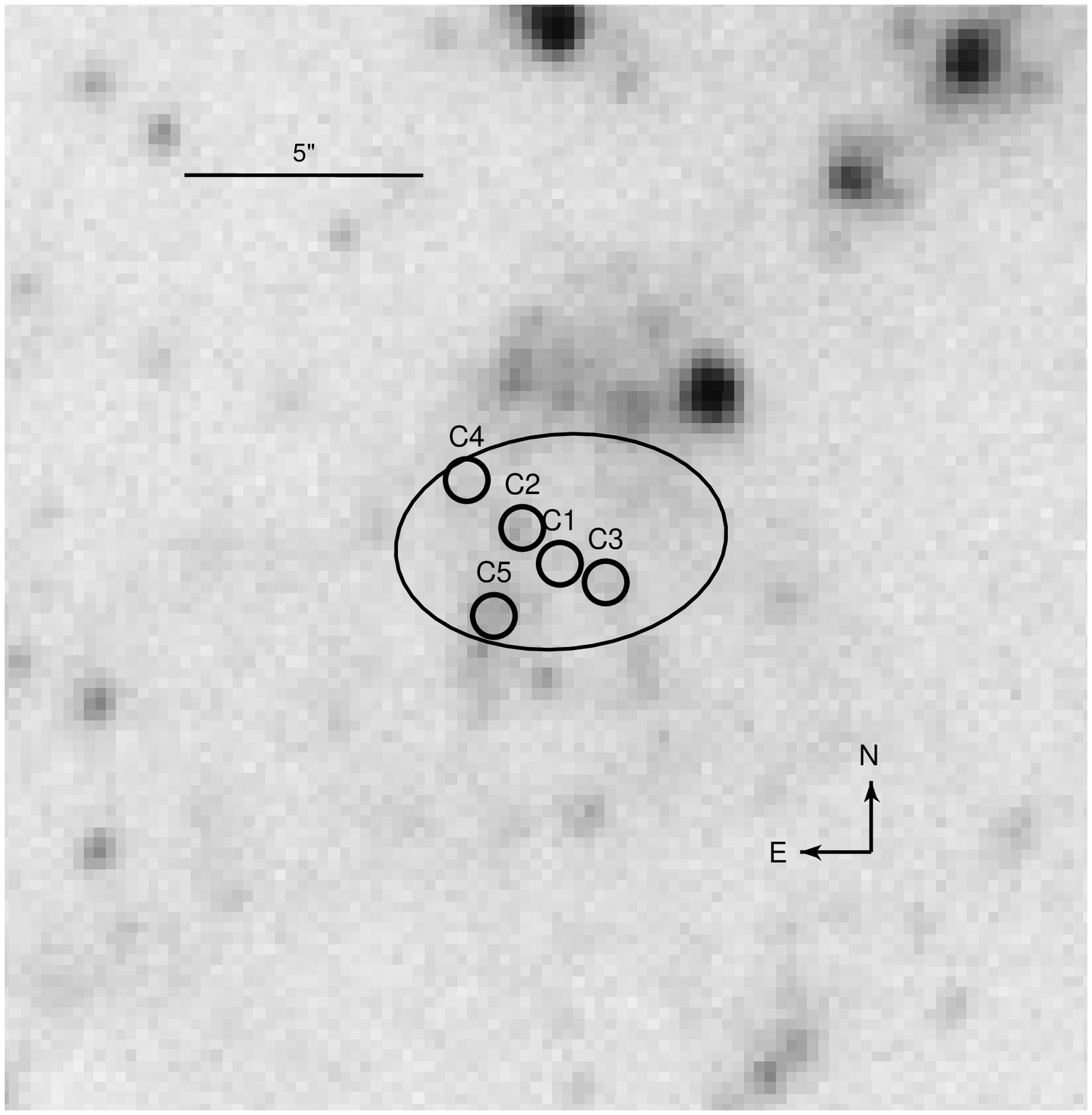}
\caption{The optical identification of the ULX in NGC628. The left panel is a V
image taken with the 6.4 meter Magellan/Baade telescope. Overlayed on the image
are the $3\sigma$ error ellipses for the X-ray positions calculated with
WAVDETECT. The ULX is on a chain of star forming knots that strays away from a
main spiral arm. The right panel shows the immediate environments of the ULX.
}

\end{figure}

\begin{figure}
\plotone{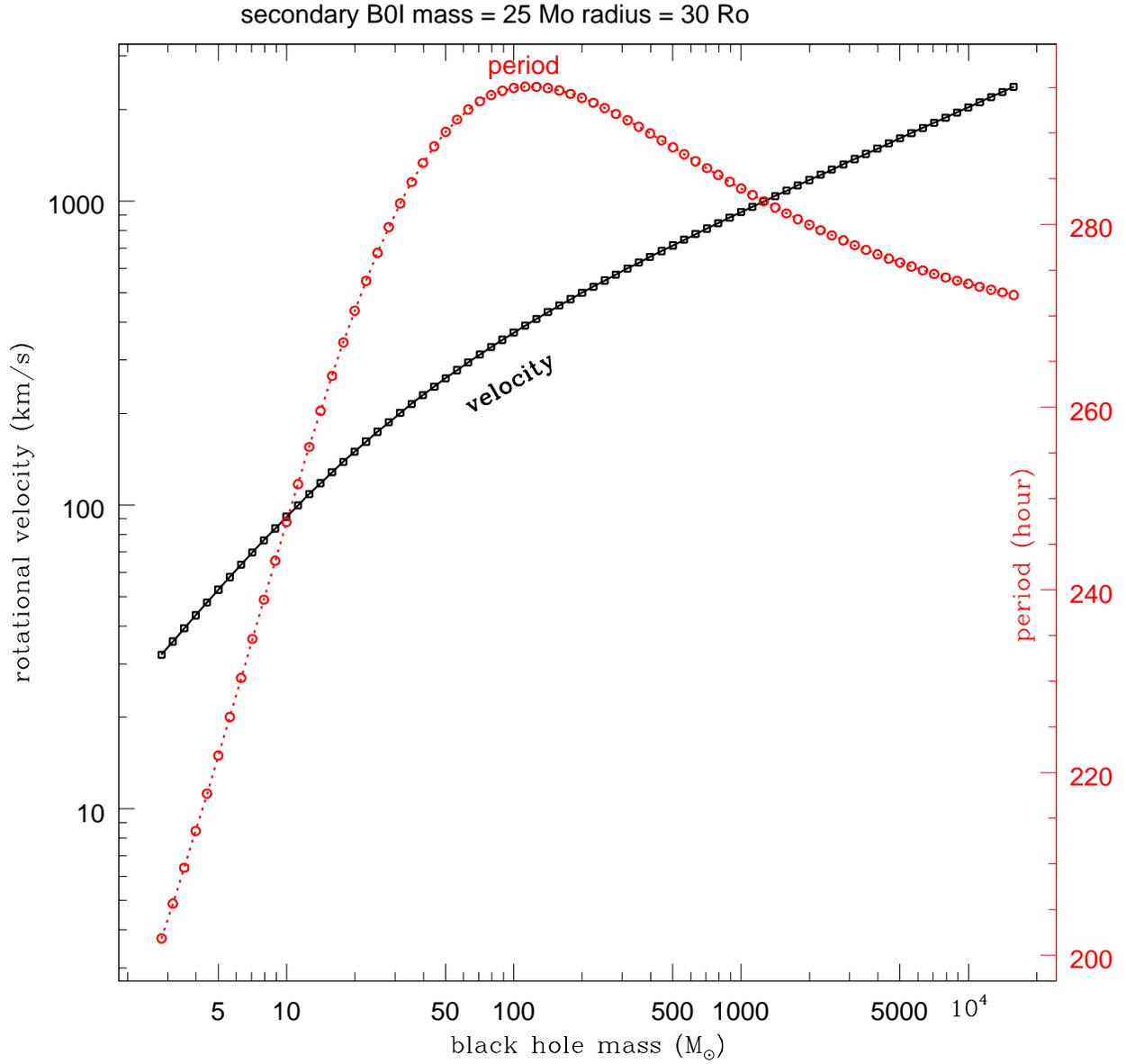}
\caption{The expected orbital period and rotational velocity for a binary
system with a B0 Ib secondary ($R=30R_\odot$,$M=25M_\odot$) filling its Roche lobe. }

\end{figure}

\end{document}